\documentclass[notitlepage,prd,twocolumn,showpacs,preprintnumbers,nofootinbib,superscriptaddress,floatfix,tightenlines]{revtex4-2}
\usepackage{mathrsfs}
\usepackage{amssymb}
\usepackage{amsmath}
\usepackage{graphicx}
\usepackage{subfig}
\usepackage[utf8]{inputenc}
\usepackage{amsfonts,amsbsy}
\usepackage{nicefrac}
\usepackage{xcolor}
\definecolor{lcolor}{rgb}{0.5,0,0}
\definecolor{citcolor}{rgb}{0,0.3,0.0}
\usepackage[breaklinks,colorlinks,urlcolor=blue,citecolor=citcolor,linkcolor=lcolor]{hyperref}
\usepackage{placeins}
\usepackage{comment}
\usepackage{bbm}

\usepackage[normalem]{ulem} % provides e.g. \sout{...} for strikeout
\usepackage{enumitem}

\newcommand{\Tr}{\mathrm{Tr}}

\newcommand{\qs}{Q_\mathrm{s}}
\newcommand{\nc}{N_\mathrm{c}}

\newcommand{\fig}{Fig.~}
\newcommand{\figs}{Figs.~}
\newcommand{\eq}{Eq.~}

\newcommand{\eqs}{Eqs.~}

\newcommand{\nr}[1]{(\ref{#1})}

\newcommand{\ud}{\mathrm{d}}

\newcommand{\xt}{\mathbf{x}}

\newcommand{\yt}{\mathbf{y}}

\newcommand{\zt}{\mathbf{z}}

\newcommand{\bt}{\mathbf{b}}

\newcommand{\kt}{\mathbf{k}}

\newcommand{\at}{\mathbf{a}}
\newcommand{\ct}{\mathbf{c}}
\newcommand{\dt}{\mathbf{d}}
\newcommand{\Bt}{\mathbf{B}}

\begin{document}

\title{Hot spots and gluon field fluctuations as causes of eccentricity in small systems}

\author{S.~Demirci} 
\affiliation{Department of Physics, P.O.~Box 35, 40014 University of Jyv\"{a}skyl\"{a}, Finland}
\affiliation{Helsinki Institute of Physics, P.O.~Box 64, 00014 University of Helsinki, Finland}

\author{T.~Lappi} 
\affiliation{Department of Physics, P.O.~Box 35, 40014 University of Jyv\"{a}skyl\"{a}, Finland}
\affiliation{Helsinki Institute of Physics, P.O.~Box 64, 00014 University of Helsinki, Finland}

\author{S.~Schlichting} 
\affiliation{Fakult\"at für Physik, Universit\"at Bielefeld, D-33615 Bielefeld, Germany}

\begin{abstract}
We calculate eccentricities in high energy proton-nucleus collisions, by calculating correlation functions of the energy density field of the Glasma immediately after the collision event at proper time \(\tau = 0^{+}\). We separately consider the effects  of color charge and geometrical hot spot fluctuations, analytically performing the averages over both in a dilute-dense limit. We show that geometric fluctuations of hot spots inside the proton are the dominant source of eccentricity whereas color charge fluctuations only give a negligible correction. The size and number of hot spots are the most important parameters characterizing the eccentricities.

\end{abstract}
\maketitle

\section{Introduction}

Collective azimuthal correlations, commonly parametrized in terms of harmonic ``flow'' coefficients $v_n$ of produced particles, were a crucial experimental signal in the discovery of the strongly interacting Quark-Gluon Plasma (QGP) in high-energy heavy ion collisions. More recently, very similar signals of collective correlations have also been discovered in high multiplicity events of various smaller collision systems, including proton-nucleus (pA) and proton-proton (pp) collisions at the Large Hadron Collider (LHC)~\cite{Khachatryan:2010gv,Adare:2013piz,Aad:2012gla,Aad:2014lta,ABELEV:2013wsa,Khachatryan:2014jra,CMS:2012qk,Abelev:2012ola,Abelev:2014mda,Khachatryan:2015waa,Aad:2013fja}, as well as proton/deuteron/helium-nucleus collisions at the Relativistic Heavy Ion Collider (RHIC)~\cite{PHENIX:2018lia}. Since the lifetime of any such smaller collision system is significantly shorter than that of a heavy-ion collision, the emergence of such correlations was initially unexpected. This has led to an intense discussion \cite{Schenke:2017bog} on whether the observed correlations in small collision system should be attributed to momentum-space correlations already present in the colliding projectiles~\cite{Kovner:2010xk,Dumitru:2010iy,Dusling:2012iga,Kovner:2011pe,Kovner:2012jm,Lappi:2015vta,Schenke:2015aqa,Lappi:2015vha}, or whether they result from the final state response to the coordinate space geometry, either via hydrodynamical evolution~\cite{Bozek:2011if,Bozek:2012gr,Werner:2013ipa} or by a simpler scattering mechanism~\cite{Xu:2011fi,Uphoff:2014cba,He:2015hfa,Koop:2015wea,Kurkela:2018qeb,Romatschke:2018wgi,Kurkela:2019kip}.

Specifically, for the final state response to the coordinate space geometry, it was quickly realized that the subnucleonic degrees of freedom are crucial for understanding azimuthal correlations in pp and pA collisions~\cite{Bzdak:2013zma,Schenke:2014zha,Albacete:2016pmp,Albacete:2016gxu,Weller:2017tsr,Mantysaari:2017cni,Moreland:2018gsh,Mantysaari:2020axf}. This has lead to a series of new investigations into the transverse spatial distribution of subnucleonic degrees of freedom in protons and nuclei \cite{Schlichting:2014ipa,Albacete:2016pmp,Mantysaari:2016ykx,Mantysaari:2016jaz,Mantysaari:2017dwh,Mantysaari:2018zdd}. Potential sources of subnucleon scale correlations in the energy density include both hot spot-like correlations, perhaps originating in the valence quark structure of the nucleon, and color charge fluctuations that ultimately result from the quantum mechanical randomness in the gluon radiation that generates the small-$x$ degrees of freedom in a proton or nucleus. While a systematical description including both of these aspects is only beginning to take shape~\cite{Dumitru:2020gla}, both aspects of subnucleon scale correlations can already be included in specific models, and the purpose of this paper is to construct a simple and transparent model for the initial energy density in a proton-nucleus collision, including some of the most important sources of fluctuations at a subnucleonic scale.

We will adopt the Color-Glass Condensate (CGC)~\cite{Gelis:2010nm} picture as a natural framework for discussing different sources of correlations and fluctuations of the initial energy density  as a function of the transverse coordinate.  We will include color charge fluctuations  in a Gaussian model~\cite{McLerran:1993ni,McLerran:1993ka,McLerran:1994vd} parametrization in both the dilute probe and the dense target. The target nucleus will be treated as homogenous on average, with the Gaussian color charge fluctuations included to all orders in the color field. The inhomogenous probe proton, on the other hand, will be treated as dilute enough to linearize the energy density in its color charge, which also fluctuates as a Gaussian local charge density as in the MV model~\cite{McLerran:1993ni,McLerran:1993ka,McLerran:1994vd}. In addition to color charge fluctuations, we will consider the probe as consisting of $N_q$ hot spots of size $r$, with locations  that fluctuate within the size of the probe $R$ on an event-by-event basis. If these hot spots are thought of as originating from valence quarks, one would take $N_q=3$ for the proton, but we will remain more agnostic and consider $N_q$ as a free parameter, as in e.g. Ref.~\cite{Moreland:2018gsh}. 

Notably, the degrees of freedom in our model are basically the same as in the IP-Glasma model~\cite{Schenke:2012wb,Schenke:2012hg,Mantysaari:2017cni}, and to a first approximation our model could be considered as an analytically tractable simplification of IP-Glasma in the dilute-dense limit. However, two significant simplifications with respect to the IP-Glasma model need to be mentioned. First of all, unlike in IP-Glasma, we do not include any nucleon-level (``MC Glauber'') structure on the nucleus side, but our target nucleus is always a homogenous (on average) and infinite sheet, corresponding to the large nuclear mass number ($A \to \infty$) limit of the IP-Glasma model for the nucleus. The fluctuating nucleon positions inside the nucleus that are neglected here would naturally be important in order to describe the eccentricities of the created system. Secondly, while the Gaussian fluctuations of the color charges introduce some event-by-event fluctuations in the total energy produced, the IP-Glasma model also features additional ``$\qs$ fluctuations'', which are not included in our model. In fact, the effect of such fluctuations is   to make the color charge fluctuations non-Gaussian~\cite{Lappi:2015vta}. In addition to the eccentricities, these two sources of fluctuations are also important to describe the fluctuations in total multiplicity, i.e. different ``centrality'' classes. In fact, describing multiplicity fluctuations in proton-proton collisions has been a part of the motivation for introducing the ``$\qs$ fluctuations''~\cite{McLerran:2015qxa,Bzdak:2015eii}.

Due to the very asymmetric treatment of the probe and target, our model might not be the best to fit experimental data, although we believe that it should be a good approximation for purely fluctuation-driven systems. Of course, in the future it could also extended by additional features such as nucleon position fluctuations in the target; however this is not the primary goal  of this study. Instead, our purpose is to construct an analytically tractable model that enables us to cleanly study the effect of different sources of fluctuations, and of the values of parameters that describe them. The extreme dilute-dense limit enables us to perform all the averages over different types of fluctuations analytically. Since color charge fluctuations dominate on the side of the target, and hot spot positions fluctuations on the side of the probe, we can cleanly distinguish the effects of the two separately. 

Our model depends on a set of phenomenological parameters that will be discussed explicitly below. These are the number of hot spots $N_q$, the size of a hot spot $r$, the size of the proton $R$, and IR regulator for the long range Coulomb tails of the color field $m$, a UV cutoff $C_0$ and the saturation scales for the proton $\tilde{\mu}$ and the nucleus $\qs$. The same parameters are also present in the IP-Glasma model (where the lattice spacing $a$ provides the UV cutoff). We will calculate initial energy densities at $\tau=0$, where they can be expressed analytically. We then obtain analytical expressions for the two-point energy density correlation function, which is the fundamental object characterizing the fluctuations and correlations of the energy density. From the energy density correlator we can then calculate eccentricities in a straightforward manner. The eccentricities are mostly driven by the long distance behavior of the energy density correlator.  In particular this means that they are, contrary to the value of the energy density,  quite insensitive to the UV cutoff. Since we use the two-point function of the energy  density, we can only get ``2-particle''  eccentricities $\varepsilon_n\{2\}$. 

We will start this paper by briefly reviewing in Sec.~\ref{sec:endens} how the initial energy density in the Glasma is calculated from the Wilson lines representing the color fields of the colliding projectiles. We will then in Sec.~\ref{sec:A} discuss how Wilson line correlators are calculated in our model for the nucleus, which is a the homogenous, infinite, fluctuating system of Gaussian color charges that is treated to all orders in the color fields. Further details on the Gaussian averaging procedure are given in Appendix~\ref{app:gaussian}. We then introduce in Sec.~\ref{sec:p} our hot spot model for the proton, and analytically perform the geometric averages over the locations of the hot spots and over the color charges in the individual hot spots, which are treated as dilute objects. We then use this model to calculate the two-point correlation function of the energy density in Sec.~\ref{sec:eps}, demonstrating the sensitivity (or insensitivity) of these quantities to the parameters in our model. We then continue in Sec.~\ref{sec:ecc} to obtain values of the $n=2,3,4$ harmonics, i.e. the eccentricities $\varepsilon_n\{2\}$.
Further checks of the dependence on the parameters are made in Appendix~\ref{app:numerics}. For comparison, some results are worked out in Appendix ~\ref{app:delta} in a model  similar in spirit to that introduced in Ref.~\cite{Blaizot:2014nia}, where the energy density is a superposition of purely pointlike hot spots, which are only correlated through the constraint that their center of mass lies at the origin.  We then conclude in Sec.~\ref{sec:conc}, pointing out future directions and discussing how our model could be further constrained in the future.

%%%%%%%%%%%%%%%%%%%%%%%%%%%%%%%%%%%%%%%%%%%%%%%%%%%%%%%%%%%%%%%%%%%%%%%
\section{Energy density in glasma}
\label{sec:endens}

In the CGC effective theory a high energy nucleon or nucleus is described as a cloud of small-$x$ gluons radiated by large-$x$ partons, represented as color charges. The gluon field is taken to be so dense that it can be described by a classical color field. The color field of a lone nucleus can then be expressed analytically as a function of its color charges. When the result for the color field is gauge rotated to the light cone gauge, one finds that the field is purely transverse and the transverse components are gauge rotations of the vacuum, sometimes referred to as ``transverse pure gauge fields''. The expressions for the color fields of two nuclei far away from each other, in their respective light cone gauges, are~\cite{McLerran:1993ka,McLerran:1994vd}
\begin{equation}
\alpha^i_{\xt} = \frac{i}{g} U_{\xt} \partial ^i U_{\xt}^{\dagger}  \quad \beta^i_{\xt} = \frac{i}{g} V_{\xt} \partial ^i V_{\xt}^{\dagger}.
\end{equation}
Here \(U_{\xt}\) and \(V_{\xt}\) are light-like Wilson lines, the relevant degrees of freedom in CGC, that describe the eikonal interaction of a color charge moving through a color field.

In a heavy ion collision one has two such color fields passing through and interacting with each other forming a non-equilibrium Glasma state \cite{Lappi:2006fp}, which continues to evolve into a Quark-Gluon Plasma. Immediately after the collision the interaction between the gluon clouds of the colliding CGC sheets leads to the creation of a longitudinal component of the gauge field. The gauge potentials  can be expressed, in Fock-Schwinger $(A^{\tau}=0)$ gauge and at proper time \(\tau={0^+}\), as \cite{Kovner:1995ja,Kovner:1995ts}
\begin{equation}
A^i_{\xt} = \alpha_{\xt}^i + \beta_{\xt}^i, \quad 
A^{\eta}=\frac{ig}{2} \big[ \alpha_{\xt}^i, \beta_{\xt}^i \big],
\label{eq:Afields0}
\end{equation}
where \(\eta\) is the space-time rapidity.

Based on the results in Eq.~(\ref{eq:Afields0}) one can compute the energy density at \(\tau=0^+\) ~\cite{Lappi:2006fp,Lappi:2017skr,Albacete:2018bbv}
\begin{equation}\label{eq:endens}
\begin{split}
\langle \varepsilon (\xt) \rangle = (-ig)^2 & (\delta^{ij}\delta^{kl}+\varepsilon^{ij}\varepsilon^{kl})\frac{1}{2}if^{abc}if^{a'b'c} \\
\times & \langle \alpha^{i,a}_{\xt} \alpha^{k,a'}_{\xt} \rangle \langle \beta^{j,b}_{\yt} \beta^{l,b'}_{\yt} \rangle
\end{split}
\end{equation}
and straightforwardly get the energy density two-point function
\begin{equation}\label{eq:endens2}
\begin{split}
\langle \varepsilon(\textbf{x})\varepsilon(\textbf{y}) \rangle &= (-ig)^4 (\delta^{ij}\delta^{kl}+\varepsilon^{ij}\varepsilon^{kl})(\delta^{i'j'}\delta^{k'l'}+\varepsilon^{i'j'}\varepsilon^{k'l'}) \\
& \times \frac{1}{4} if^{abe}if^{cde}if^{a'b'e'}if^{c'd'e'} \\
& \times \big \langle \alpha^{i,a}_{\textbf{x}} \alpha^{k,c}_{\textbf{x}} \alpha^{i',a'}_{\textbf{y}} \alpha^{k',c'}_{\textbf{y}} \big \rangle
\big \langle \beta^{j,b}_{\textbf{x}} \beta^{l,d}_{\textbf{x}} \beta^{j',b'}_{\textbf{y}} \beta^{l',d'}_{\textbf{y}} \big \rangle.
\end{split}
\end{equation}
Here we have expanded the gluon fields in the Lie algebra
\begin{equation} \label{eq:alphaDecomp}
\begin{split}
&
\alpha _{\textbf{x}} ^{i} = \frac{i}{g} U_{\textbf{x}} \partial ^i U ^{\dagger} _{\textbf{x}}, 
\qquad \quad 
\alpha ^i _{\textbf{x}} = \alpha ^{i,a} _{\textbf{x}} t^a, 
\\ &
\alpha ^{i,a} _{\textbf{x}} = \frac{2i}{g} \Tr[t^a U_{\textbf{x}} \partial ^i U ^{\dagger} _{\textbf{x}}].
\end{split}
\end{equation}
The brackets $\langle .\rangle$ in \eqs\nr{eq:endens} and~\nr{eq:endens2} refer to an averaging over the color charges, or distributions of Wilson lines, in the two colliding nuclei separately. Since the color fields of the nuclei are built up slowly a long time before the collision, the two nuclei are not correlated. Thus the expectation values for the colliding nuclei factorize from each other  and
can be computed separately. In this work, we take the \(\alpha\)-part to describe a dense nucleus and the \(\beta\)-part to describe a dilute proton.

We will also separate the energy density two-point function into distinct connected and disconnected parts. We introduce the following decomposition of the four $\alpha$ and $\beta$ correlators:
\begin{multline}
\label{disc-conn}
\left< \alpha^{i,a}_{\textbf{x}} \alpha^{k,c}_{\textbf{x}} \alpha^{i',a'}_{\textbf{y}} \alpha^{k',c'}_{\textbf{y}} \right> =  \\  
 \left< \vphantom{\alpha^{k',c'}_{\textbf{y}}}
 \alpha^{i,a}_{\textbf{x}} \alpha^{k,c}_{\textbf{x}} \right> \left< \alpha^{i',a'}_{\textbf{y}} \alpha^{k',c'}_{\textbf{y}} \right> + \left< \alpha^{i,a}_{\textbf{x}} \alpha^{k,c}_{\textbf{x}} \alpha^{i',a'}_{\textbf{y}} \alpha^{k',c'}_{\textbf{y}} \right>_{C}
\end{multline}
where disconnected contributions $\big \langle \alpha^{i,a}_{\textbf{x}} \alpha^{k,c}_{\textbf{x}} \big \rangle \big \langle \alpha^{i',a'}_{\textbf{y}} \alpha^{k',c'}_{\textbf{y}} \big \rangle$ refer to the parts appearing in expectation value of the energy density, while the connected part $\big \langle \alpha^{i,a}_{\textbf{x}} \alpha^{k,c}_{\textbf{x}} \alpha^{i',a'}_{\textbf{y}} \alpha^{k',c'}_{\textbf{y}} \big \rangle_{C}$ of a correlator of $\alpha$'s or $\beta$'s  refers to the full correlator minus the disconnected part. The disconnected part of the energy density two-point function is just the product of energy density expectation values, given by the product of the disconnected parts of the $\alpha$- and $\beta$-correlators. We refer to the contribution $\big \langle \alpha^{i,a}_{\textbf{x}} \alpha^{k,c}_{\textbf{x}} \big \rangle \big \langle \alpha^{i',a'}_{\textbf{y}} \alpha^{k',c'}_{\textbf{y}} \big \rangle
\big \langle \beta^{j,b}_{\textbf{x}} \beta^{l,d}_{\textbf{x}} \beta^{j',b'}_{\textbf{y}} \beta^{l',d'}_{\textbf{y}} \big \rangle_{C}$ from the disconnected part of the nucleus ($\alpha$)  correlator and the connected part of the proton ($\beta$) correlator as the ``proton fluctuation'' part of the energy density correlator. Analogously the connected-nucleus, disconnected-proton contribution $\big \langle \alpha^{i,a}_{\textbf{x}} \alpha^{k,c}_{\textbf{x}} \alpha^{i',a'}_{\textbf{y}} \alpha^{k',c'}_{\textbf{y}} \big \rangle_{C} \big \langle \beta^{j,b}_{\textbf{x}} \beta^{l,d}_{\textbf{x}} \big \rangle \big \langle \beta^{j',b'}_{\textbf{y}} \beta^{l',d'}_{\textbf{y}} \big \rangle$ is referred to as the ``nucleus fluctuation'' part of the energy density two-point function. We expect that in our approximation the fully connected  
contribution (on both proton and nucleus sides)
$\big \langle \alpha^{i,a}_{\textbf{x}} \alpha^{k,c}_{\textbf{x}} \alpha^{i',a'}_{\textbf{y}} \alpha^{k',c'}_{\textbf{y}} \big \rangle_{C}
\big \langle \beta^{j,b}_{\textbf{x}} \beta^{l,d}_{\textbf{x}} \beta^{j',b'}_{\textbf{y}} \beta^{l',d'}_{\textbf{y}} \big \rangle_{C}$
only gives a small contribution as it is sensitive to the fluctuations of both the proton and the nucleus, and we will neglect it in our calculation of the energy density two-point function in Eq.~(\ref{eq:endens2}).

We will now go on to compute the 4-\(\alpha\) correlator for the nucleus in the Sec.~\ref{sec:A} and the 4-\(\beta\) correlator for the proton in the Sec.~\ref{sec:p}. The nucleus part will be computed as a full nonlinear Gaussian and the proton will be expanded to the first order in the proton saturation scale in the dilute limit.

%%%%%%%%%%%%%%%%%%%%%%%%%%%%%%%%%%%%%%%%%%%%%%%%%%%%%%%%%%%%%%%%%%%%%%%%%%%%%
\section{Nucleus: fundamental representation 8-point correlators}
\label{sec:A}

Now we will compute the 4-point function of the nucleus gluon fields \(\alpha\) as a nonlinear Gaussian expectation value, where we use the term ``Gaussian'' in the sense of Gaussian contractions of color charge densities. The calculation yields a result in terms of a generic two-point function of Wilson lines. We will then for simplicity adopt for the nucleus the GBW~\cite{GolecBiernat:1998js} parametrization for this two-point function. The GBW form is Gaussian in another sense, namely that it assumes a Gaussian functional form for the Wilson line dipole expectation value as a function of the transverse coordinate separation
\begin{equation}
    \left< \frac{1}{\nc}\Tr U^\dag_{\xt} U_{\yt}\right> = \exp\left[-\qs^2(\xt-\yt)^2/4 \right].
\end{equation}
The GBW parametrization yields slightly simpler final expressions, since the crossed partial derivatives $\partial^1_{\xt}\partial^2_{\xt}$ of the pure second power $(\xt-\yt)^2$ in the exponent vanish\footnote{In other words the linearly polarized gluon distribution vanishes~\cite{Lappi:2017skr}.}. The algorithm for calculating Wilson line expectation values with Gaussian color charges is discussed in more detail in Appendix~\ref{app:gaussian}. Here the nuclear saturation scale $\qs$ is the only parameter needed to characterize our infinite, homogenous target nucleus.

We start by expanding the \(\alpha\)'s using their definition in \eq\eqref{eq:alphaDecomp}, yielding
\begin{multline}
 \langle \alpha^{i,a}_{\xt} \alpha^{k,c}_{\xt} \alpha^{i',a'}_{\yt} \alpha^{k',c'}_{\yt} \rangle
\\
= 
\frac{16}{g^4}
\langle  
 \Tr[t^a U_{\xt} \partial^i_{\xt} U^{\dagger}_{\xt}] 
 \Tr[t^c U_{\xt} \partial^k_{\xt} U^{\dagger}_{\xt}] 
 \\  \times
 \Tr[t^{a'} U_{\yt} \partial^{i'}_{\yt} U^{\dagger}_{\yt}] 
 \Tr[t^{c'} U_{\yt} \partial^{k'}_{\yt} U^{\dagger}_{\yt}] 
\rangle.
\end{multline}
Next we introduce new transverse coordinates to be able to pull out the derivatives and we write the traces in index notation, such that
\begin{equation}
\begin{split}
& \langle \alpha^{i,a}_{\xt} \alpha^{k,c}_{\xt} \alpha^{i',a'}_{\yt} \alpha^{k',c'}_{\yt} \rangle
\\
&=
\lim_{\substack{x_i \to x \\ y_i \to y}} \Bigg\{
\frac{16}{g^4}
\partial^i_{\xt_2} \partial^k_{\xt_4} \partial^{i'}_{\yt_2} \partial^{k'}_{\yt_4}
t^a_{k_2 k_1} t^c_{k_4 k_3} t^{a'}_{k_6 k_5} t^{c'}_{k_8 k_7}
\\ & \times
\langle  
  [U_{\xt_1} U^{\dagger}_{\xt_2}]_{k_1 k_2} 
  [U_{\xt_3} U^{\dagger}_{\xt_4}]_{k_3 k_4} 
  [U_{\yt_1} U^{\dagger}_{\yt_2}]_{k_5 k_6} 
  [U_{\yt_3} U^{\dagger}_{\yt_4}]_{k_7 k_8} 
\rangle \Bigg\}.
\end{split}
\end{equation}
This 8-point function of Wilson lines can be computed, using the algorithm used in \cite{Blaizot:2004wv}, as
\begin{multline}
 \langle \alpha^{i,a}_{\xt} \alpha^{k,c}_{\xt} \alpha^{i',a'}_{\yt} \alpha^{k',c'}_{\yt} \rangle
 =
\lim_{\substack{x_i \to x \\ y_i \to y}}
\left\{
\vphantom{
  \begin{bmatrix} t^a_{k_2 k_1} t^c_{k_4 k_3} t^{a'}_{k_6 k_5} t^{c'}_{k_8 k_7} \\ 0 \\ \vdots \\ 0 \end{bmatrix}
}
\frac{16}{g^4}
\partial^i_{\xt_2} \partial^k_{\xt_4} \partial^{i'}_{\yt_2} \partial^{k'}_{\yt_4}
\right.
\\  \times
\left.
\begin{bmatrix}
\delta^{k_1 k_2} \delta^{k_3 k_4}  \delta^{k_5 k_6} \delta^{k_7 k_8} \\ 
\delta^{k_2 k_7} \delta^{k_1 k_8}  \delta^{k_3 k_4} \delta^{k_5 k_6} \\ 
\vdots
\end{bmatrix}
^{\mathrm{T}}
\!\!
e^{ M_{24\times24}}
\!\!
\begin{bmatrix}
t^a_{k_2 k_1} t^c_{k_4 k_3} t^{a'}_{k_6 k_5} t^{c'}_{k_8 k_7} \\
0 \\
\vdots \\
0
\end{bmatrix}
\right\}.
\end{multline}
Here the 24$\times$24 matrix $M_{24\times24}$ depends on the two-point correlator of Wilson lines at the coordinates $\xt_i, \yt_i$ and color factors, and can be obtained as discussed in  Appendix~\ref{app:gaussian}. We will not write the rather lengthy expression here. The color structures $\delta^{a_1 a_2} \delta^{a_3 a_4}  \delta^{a_5 a_6} \delta^{a_7 a_8}$ etc. on the left are the elements of the chosen basis for different singlet operators that can be formed from the 8 Wilson lines, and the structure $t^a_{a_2 a_1} t^c_{a_4 a_3} t^{a'}_{a_6 a_5} t^{c'}_{a_8 a_7} $ on the right corresponds to the particular correlator that we need to calculate here. 

Instead of attempting to explicitly exponentiate the full 24$\times$24 matrix, it is useful to first take the derivatives. We use the following identity for the derivative of a matrix exponential 
\begin{equation}
\partial_x e^{M(x)} = \int_0^1 \ud t e^{tM}[\partial_x M]e^{(1-t)M} ,
\end{equation}
where \(M \equiv M_{n \times n}\). As one can see, the derivatives do not enter the matrix exponentials themselves. Thus after taking the derivatives, we can take the coordinate limits $\xt_i\to \xt, \yt_i\to \yt$ in the matrix exponentials, rendering their evaluation easier. 
The result is not written here for it is a long equation and is not a new result. It was computed with a slightly different formulation  of the same algorithm  in Ref.~\cite{Albacete:2018bbv} and it corresponds to the result we get using this method. 

%%%%%%%%%%%%%%%%%%%%%%%%%%%%%%%%%%%%%%%%%%%%%%%%%%%%
\section{Proton: hot spot model}
\label{sec:p}

We will treat the proton as a collection of Gaussian hot spots. The hot spots consist of color charges with Gaussian fluctuations that are local (uncorrelated between different points), i.e. given by the MV model~\cite{McLerran:1993ni,McLerran:1993ka,McLerran:1994vd}. The hot spot approach is inspired by the picture of a proton having three large-$x$ valence quarks and has been used in the IP-Glasma model~\cite{Schlichting:2014ipa,Mantysaari:2016ykx}. We also let these hot spots to be Gaussianly distributed in the transverse coordinate inside the proton with respect to its center of mass.

We now have two averages to calculate in order to obtain the energy density correlator. We define a double average of an operator
\begin{multline} \label{eq:defavg}
\langle \langle \mathcal{O} \rangle \rangle =
\Big( \frac{2 \pi R^2}{N_q} \Big)
\int \prod^{N_q}_{i=1} \Big[ \ud^2\bt_i T(\bt_i-\mathbf{B})\Big]
\\  \times
\delta^{(2)}\left(
\frac{1}{N_q}\sum _{i=1}^{N_q}\bt_i - \mathbf{B}\right)  \langle \mathcal{O} \rangle _{CGC}
\end{multline}
to include both the average over the color sources in the MV-model, and the averaging over the hot spot coordinates.
Since we eventually want to calculate eccentricities with respect to the center of proton, which fluctuates with the hot spot positions, it is essential to  explicitly fix the center of mass of the hot spot system\footnote{This is similar to what is done in practice in Monte Carlo simulations, where one first generates the initial configurations, and then uses them to calculate the center of the system. In our approach we fix the center of mass of the hot spot coordinates to $\mathbf{B}$, but do not take into account the fluctuations of the color charges in the determination of the center of mass. The latter is a much smaller effect and would be difficult to incorporate in our analytical approach.} to a known coordinate $\mathbf{B}$.
The distribution of the hot spot locations is taken to be a Gaussian:
\begin{equation}\label{eq:pprofile}
T(\bt) = \frac{1}{2 \pi R^2} \exp\left[-\frac{\bt ^2}{2 R^2}\right],
\end{equation}
where the parameter \(R\) has an interpretation as the proton radius. The prefactor in \eq\nr{eq:defavg} is chosen so that the expectation value is normalized: $ \langle \langle 1 \rangle \rangle =1 $.

The distribution of pointlike color  charges within a hot spot is taken to have a Gaussian distribution
\begin{equation}\label{eq:hsprofile}
\mu^2(\xt)=\frac{\mu^2_0}{2 \pi r^2} \exp \left[-\frac{\xt ^2}{2 r^2}\right],
\end{equation}
where \(r\) is the radius of the hot spot and \(\mu _0\) is a coefficient characterizing the color charge density. The hot spots enter the calculation through the two-point function of color charge densities as follows
\begin{equation}\label{eq:p2point}
\big \langle \rho ^a(\xt) \rho ^b(\yt) \big \rangle = \sum_{i=1}^{N_q} \mu ^2 \Big(\frac{\xt+\yt}{2} - \bt _i\Big) \delta^{(2)}(\xt - \yt) \delta^{ab},
\end{equation}
where \(N_q\) is the number of hot spots in the proton. Correlators of more than two color charge densities are calculated by taking the distribution to be Gaussian, and can thus be expressed in terms of the two point function~\nr{eq:p2point}.

Now we can compute the proton contribution to the energy density two-point function.
\begin{equation}
\langle \langle \beta^{j,b}_{\xt} \beta^{l,d}_{\xt} \beta^{j',b'}_{\yt} \beta^{l',d'}_{\yt} \rangle \rangle,
\end{equation}
which we will do in the limit of small charge density, i.e. to lowest order in the parameter $\mu_0$.
We start by writing the color fields \(\beta\) in terms of the Wilson lines
\begin{multline}
\big\langle\big\langle\beta^{j,b}_{\xt} \beta^{l,d}_{\xt} \beta^{j',b'}_{\yt} \beta^{l',d'}_{\yt} \big\rangle\big\rangle
\\
=  \frac{16}{g^4}
\big\langle\big\langle
 \Tr[t^b V_{\xt} \partial^j_{\xt} V^{\dagger}_{\xt}] 
 \Tr[t^d V_{\xt} \partial^l_{\xt} V^{\dagger}_{\xt}] 
 \\ \times
 \Tr[t^{b'} V_{\yt} \partial^{j'}_{\yt} V^{\dagger}_{\yt}] 
 \Tr[t^{d'} V_{\yt} \partial^{l'}_{\yt} V^{\dagger}_{\yt}] 
\big\rangle\big\rangle.
\end{multline}
We then expand to lowest order in the color sources yielding
\begin{multline}
\big\langle \big\langle \beta^{j,b}_{\xt} \beta^{l,d}_{\xt} \beta^{j',b'}_{\yt} \beta^{l',d'}_{\yt} \big\rangle \big\rangle 
=
\\
\int 
\ud^2\at \, \ud^2\bt \, \ud^2\ct \, \ud^2\dt 
\,
% \\  \times
\textbf{G}^j_{\xt}(\xt-\at)  
\textbf{G}^l_{\xt}(\xt-\bt)
\\ \times
\textbf{G}^{j'}_{\yt}(\yt-\ct)
\textbf{G}^{l'}_{\yt}(\yt-\dt)
\\  \times
\left<\left<  \rho^b(\at) \rho^d(\bt) \rho^{b'}(\ct) \rho^{d'}(\dt) 
\right>\right>.
\end{multline}
Here \(\textbf{G}^i_{\xt}(\xt-\yt)\) is the partial derivative in the $i$-direction of the Green's function
\begin{equation} \label{eq:GreensFunction}
G(\xt-\yt)=\int \frac{\ud^2 \kt}{(2 \pi)^2} \frac{\exp{[i \kt \cdot (\xt-\yt)]}}{\kt ^2 + m^2}, 
\end{equation}
which relates the color field in the Wilson line to the color charge density.
We  have regularized the infrared behavior of the Green's function with a mass \(m\), which should be thought of as a confinement  scale regulator \(m\sim \Lambda_{\mathrm{QCD}}\).
The derivative \(\textbf{G}^i_{\xt}(\xt-\yt)\), can, assuming \(\xt \neq \yt\),  be written as
\begin{equation}
\begin{split}
\partial ^i _{\xt} G(\xt-\yt) & \equiv \textbf{G}^i_{\xt}(\xt-\yt) 
\\ &
= -\frac{1}{2\pi} m|\xt-\yt|K_1(m|\xt-\yt|) \frac{(\xt-\yt)^i}{(\xt-\yt)^2},
\end{split}
\end{equation}
where \(K_1\) is the modified Bessel function of the second kind. This expression contains an ultraviolet, short-distance singularity at points with \(\xt = \yt\), which arises from the fact that our color charges are treated as a point-like objects. We regularize this by a short-distance cutoff, making the substitution
\begin{equation}\label{eq:defC0}
\textbf{G}^i_{\xt}(\xt-\yt) 
\rightarrow
\textbf{G}^i_{\xt}(\xt-\yt) \Theta(|\xt-\yt|-C_0)
\end{equation}
where \(\Theta\) is the Heaviside step function and $C_0$ is a short distance cutoff.

Now the color charge and hot spot averages can be evaluated fully, giving the correlator as
\begin{equation}
\label{eq:beta4Exp}
\begin{split}
&
\langle \langle \beta^{j,b}_{\xt} \beta^{l,d}_{\xt} \beta^{j',b'}_{\yt} \beta^{l',d'}_{\yt} \rangle \rangle
=
\\
&
 \int \ud^2\at \ud^2\bt \Big[ N_q F_2(\at, \bt, \Bt) + N_q(N_q-1)F_3(\at, \bt, \Bt) \Big]
\\ & \quad 
\times
\Big[
\textbf{G}^j_{\xt}(\xt-\at)
\textbf{G}^l_{\xt}(\xt-\at)
\textbf{G}^{j'}_{\yt}(\yt-\bt)
\textbf{G}^{l'}_{\yt}(\yt-\bt)
\delta^{bd}\delta^{b' d'}
\\
&\quad \quad
+ 
\textbf{G}^j_{\xt}(\xt-\at)
\textbf{G}^l_{\xt}(\xt-\bt)
\textbf{G}^{j'}_{\yt}(\yt-\at)
\textbf{G}^{l'}_{\yt}(\yt-\bt) \delta^{bb'}\delta^{dd'}
\\
&\quad \quad 
+ 
\textbf{G}^j_{\xt}(\xt-\at)
\textbf{G}^l_{\xt}(\xt-\bt)
\textbf{G}^{j'}_{\yt}(\yt-\bt)
\textbf{G}^{l'}_{\yt}(\yt-\at) \delta^{bd'}\delta^{db'}
\Big].
\end{split}
\end{equation}
This is explicitly factorized into a part describing the coordinates of the color charges $\at,\bt$, and the Green's function part describing the color field generated by these charges. In Eq.~(\ref{eq:beta4Exp}) the averages over the color charges have naturally split into two kinds of contributions. The first one, proportional to the number of hot spots $N_q$, results from taking all color charges from the same hot spot, resulting in the function
\begin{multline} \label{eq:F2}
F_2(\at, \bt, \Bt) \equiv \langle \mu^2(\at-\bt_i) \mu^2(\bt-\bt_i) \rangle_{\mathrm{Hotspot}}
\\ 
=
\left( \frac{\mu_0^2}{2\pi r^2} \right)^2 \left( \frac{1}{1+2\left(\frac{N_q-1}{N_q}\right)\frac{R^2}{r^2}} \right)
\\  \times
\exp \left\{
- 
\frac{(\at+\bt-2\Bt)^2}{4r^2 \left(1+2\left(\frac{N_q-1}{N_q}\right)\frac{R^2}{r^2}\right)}
- \frac{(\at-\bt)^2}{4r^2}
 \right\},
\end{multline}
where the second term in the exponent makes it manifest that the color charge coordinates $\at,\bt$ are within a distance $\sim r$ from each other. 
The second contribution results from taking color charges from two separate hot spots, and is proportional to $N_q(N_q-1)$, the number of pairs of distinct hot spots. It is given by the function
\begin{multline} \label{eq:F3}
F_3(\at, \bt, \Bt) \equiv \langle \mu^2(\at-\bt_i) \mu^2(\bt-\bt_j) \rangle_{\mathrm{Hotspot}}
\\ 
=
\left( \frac{\mu_0^4}{(2 \pi)^2 (R^2+r^2)} \right) \left( \frac{1}{r^2+\left(\frac{N_q-2}{N_q}\right)R^2} \right)
\\  \times
\exp \left\{ 
-
\frac{(\at+\bt-2\Bt)^2}{4\left(r^2 + \left(\frac{N_q-2}{N_q}\right)R^2\right)}
- 
\frac{(\at-\bt)^2}{4(R^2+r^2)}
\right\}
\end{multline}
where \(i,j \in \{1,\ldots,N_q\}\) and \(i\neq j.\) Here it is also clear that the coordinates of the color charges $\at,\bt$ are typically separated by  a distance $\sim R$ of the order of the size of the proton. 

Let us finally summarize here the phenomenological parameters characterizing our hot spot model. They are the proton radius parameter $R$ introduced in \eq\nr{eq:pprofile}, the proton color charge density~$\mu_0$ and hot spot size $r$ introduced in \eq\nr{eq:hsprofile}, the number of hot spots~$N_q$ introduced in \eqs\nr{eq:defavg},~\nr{eq:p2point}, the ``gluon mass'' infrared regulator for the Coulomb tails of the color field~$m$ in \eq\nr{eq:GreensFunction}, and the short distance cutoff $C_0$ in \eq\nr{eq:defC0}. We will discuss the dependence of the energy density correlator and the eccentricities on these parameters in the following sections.

%%%%%%%%%%%%%%%%%%%%%%%%%%%%%%%%%%%%%%%%%%%
\section{Results: energy density and its correlator}
\label{sec:eps}

Let us start by evaluating the energy density expectation value~\nr{eq:endens} in the framework of a full nonlinear Gaussian nucleus and hot spot model proton.
Let us first compute the part with the nucleus side average. We start by using the definition of the \(\alpha\)'s and writing the resulting expression in index notation. We will also introduce new transverse coordinates to enable us to pull the derivatives out of the correlator. Doing this we get
\begin{multline}
\langle \alpha^{i,a}_{\xt} \alpha^{k,a'}_{\xt} \rangle 
\\ 
= \lim_{\substack{\xt_i \to \xt}} \left\{ -\frac{4}{g^2} \partial^i_{\xt_2} \partial^k_{\xt_4} t^{a}_{k_2 k_1} t^{a'}_{k_4 k_3} \langle [U_{\xt_1}U^{\dagger}_{\xt_2}]_{k_1 k_2} [U_{\xt_3}U^{\dagger}_{\xt_4}]_{k_3 k_4} \rangle \right\}.
\end{multline}
Again, using the algorithm presented in \cite{Blaizot:2004wv} and discussed in more detail in Appendix~\ref{app:gaussian}, we can express this as
\begin{multline}
\langle \alpha^{i,a}_{\xt} \alpha^{k,a'}_{\xt} \rangle
\\ 
=
\lim_{\substack{\xt_i \to \xt}}
\Bigg\{
-\frac{4}{g^2} \partial ^i_{\textbf{x}_2} \partial ^k_{\textbf{x}_4}
\begin{bmatrix}
 \delta^{k_1 k_2}\delta^{k_3 k_4} \\
 \delta^{k_1 k_4}\delta^{k_2 k_3} 
\end{bmatrix}
^{\mathrm{T}}
e^{ M_{2\times2}}
\begin{bmatrix}
 t^{a}_{k_2 k_1} t^{a'}_{k_4 k_3}  \\
0
\end{bmatrix}
\Bigg\}.
\end{multline}
This reduces to
\begin{multline}
\langle \alpha^{i,a}_{\xt} \alpha^{k,a'}_{\xt} \rangle
\\ 
=
\lim_{\substack{\xt_i \to \xt}}
\Bigg\{
-\frac{2}{g^2} \delta^{aa'} \partial ^i_{\textbf{x}_2} \partial ^k_{\textbf{x}_4}
\begin{bmatrix}
0 & 1
\end{bmatrix}
e^{ M_{2\times2}}
\begin{bmatrix}
1 \\
0
\end{bmatrix}
\Bigg\}.
\end{multline}
Now we can take the derivatives of the matrix exponential and then take the coordinate limits in the similar fashion as in section \ref{sec:A}. Doing this, and using the GBW model as an input, one finds
\begin{equation}
\left< \alpha^{i,a}_{\xt} \alpha^{k,a'}_{\xt} \right> = \frac{Q_s^2 \delta^{aa'} \delta^{ik}}{2 g^2 C_F}.
\end{equation}
Plugging this back to the expression of the energy density, \eq\eqref{eq:endens}, and doing the color and transverse index algebra, we get
\begin{equation}
\langle \varepsilon(\xt) \rangle
=
\frac{N Q_s^2}{2 C_F} 
 \big \langle \beta^{j,b}_{\textbf{x}} \beta^{j,b}_{\textbf{x}} \big \rangle,
\end{equation}
where $N$ is the number of colors.

Now we will calculate the proton side contribution in the hot spot model in the dilute limit. We will start by using the definition the \(\beta\)'s and expanding the Wilson lines to the lowest order in sources yielding
\begin{multline}
\langle \varepsilon(\xt) \rangle
= \frac{N Q_s^2}{2 C_F} 
\left<\left<
\int \ud^2\yt \ud^2\zt \, \textbf{G}^j_{\xt}(\xt - \yt) \textbf{G}^j_{\xt}(\xt - \zt) 
\right.\right.
\\
\left.\left. \vphantom{\int}
\times \rho_a(\yt) \rho_a(\zt)
\right>\right>
\end{multline}
Taking the CGC and hot spot averages, we get
\begin{equation}
\label{eq:onepoint}
\begin{split}
\langle \varepsilon(\xt) \rangle 
=
&
\frac{N (N^2-1) N_q Q_s^2}{2 C_F} 
\\ & \times
\int \ud^2 \zt G^j_{\xt}(\xt - \zt) G^j_{\xt}(\xt - \zt)
F_1(\zt, \Bt),
\end{split}
\end{equation}
where we defined, analogously to \eqs\nr{eq:F2} and~\nr{eq:F3},
\begin{multline}
F_1(\zt, \Bt) \equiv \langle \mu^2(\zt-\bt_i) \rangle_{\mathrm{Hotspot}}
\\  
=
\left( \frac{\mu^2_0}{2 \pi r^2} \right)\left( \frac{1}{1+\left(\frac{N_q-1}{N_q}\right)\frac{R^2}{r^2}} \right)
\\  \times
\exp \left\{ -\frac{1}{2}  \frac{(\zt-\Bt)^2}{r^2+\left(\frac{N_q-1}{N_q}\right)R^2} \right\}
\end{multline}
with \(i \in \{1,\ldots,N_q\}\). The function $F_1$ can be interpreted as the average density of proton color charges in the transverse plane.

Now that we have obtained the expression~\nr{eq:onepoint} for the one-point function of the energy density, we move to calculating the two point function, i.e. the energy density correlator. 
We want to separate the contributions due to the proton and nucleus fluctuations. To do this, we consider separately the disconnected and connected part of the proton and nucleus contributions as defined in Eq.~(\ref{disc-conn}). The disconnected-disconnected contribution of the energy density two-point function is just the product of two energy densities in the two different transverse positions
\begin{equation}
\langle \varepsilon(\textbf{x})\varepsilon(\textbf{y}) \rangle _{\mathrm{DC,DC}} = \langle \varepsilon(\textbf{x}) \rangle \langle \varepsilon(\textbf{y}) \rangle.
\end{equation}
Note that this expression is radially symmetric in both transverse coordinates and can not have an explicit correlation between the two coordinates.

Next we want to compute the nucleus disconnected and proton connected part, i.e. the ``proton fluctuation'' contribution. Subtracting the disconnected-disconnected contribution does not lead to a particularly simple expression (see more detailed discussion in Appendix~\ref{app:numerics}), so we write this part here as
\begin{equation}
\label{eq:protonside}
\begin{split}
&
\langle \varepsilon(\textbf{x})\varepsilon(\textbf{y}) \rangle _{\mathrm{DC,C}} 
\\
&
=
\frac{Q_s^4 N^2}{4 C_F^2}
 \int \ud^2 \at \ud^2 \bt \Big[ N_q F_2(\at, \bt, \Bt) + N_q(N_q-1)F_3(\at, \bt, \Bt) \Big]
\\
&
\times
\Bigg \{
(N^2-1)^2 
\textbf{G}^i_{\xt}(\xt-\at)
\textbf{G}^i_{\xt}(\xt-\at)
\textbf{G}^j_{\yt}(\yt-\bt)
\textbf{G}^j_{\yt}(\yt-\bt)
\\ &
\quad \quad + 2(N^2-1)
\textbf{G}^i_{\xt}(\xt-\at)
\textbf{G}^i_{\xt}(\xt-\bt)
\textbf{G}^j_{\yt}(\yt-\at)
\textbf{G}^j_{\yt}(\yt-\bt)
\Bigg \}
\\& 
\quad \quad \quad - \langle \varepsilon(\textbf{x})\varepsilon(\textbf{y}) \rangle _{\mathrm{DC,DC}},
\end{split}
\end{equation}
with the same \(F_2\) and \(F_3\) as in equations \eqref{eq:F2} and \eqref{eq:F3}. Lastly, we need the nucleus connected and proton disconnected or the ``nucleus fluctuation'' contribution, which takes the form
\begin{equation}
\begin{split}
&
\langle \varepsilon(\textbf{x})\varepsilon(\textbf{y}) \rangle _{\mathrm{C,DC}} 
\\
&
= \frac{N^2 N_q^2}{(\xt - \yt)^4} \Bigg\{ \frac{8(N^2-1)}{N^2}\exp\left[-\frac{N^2 Q_s^2 (\xt - \yt)^2}{2(N^2-1)}\right] 
\\
&\quad \quad 
+N^2 Q_s^4 (\xt - \yt)^4 +4 Q_s^2 (\xt - \yt)^2 +\frac{8(1-N^2)}{N^2} \Bigg\}
\\
&\quad 
\times
\int \ud^2 \at \ud^2 \bt 
\textbf{G}^i_{\xt}(\xt-\at)
\textbf{G}^i_{\xt}(\xt-\at)
\textbf{G}^j_{\xt}(\yt-\bt)
\textbf{G}^j_{\xt}(\yt-\bt)
\\ &
\quad \quad 
\times
F_1(\at,\Bt)F_1(\bt,\Bt)
\\ & 
\quad \quad \quad - \langle \varepsilon(\textbf{x})\varepsilon(\textbf{y}) \rangle _{\mathrm{DC,DC}}
.
\end{split}
\end{equation}
As discussed earlier, we will neglect the connected-connected contribution here. Thus we now have the results for the energy density correlator that we will need.

\begin{figure*}[ptbh!]
    \centering
    \subfloat[\(N_q=1\)]{
    \includegraphics[width=1.1\columnwidth]{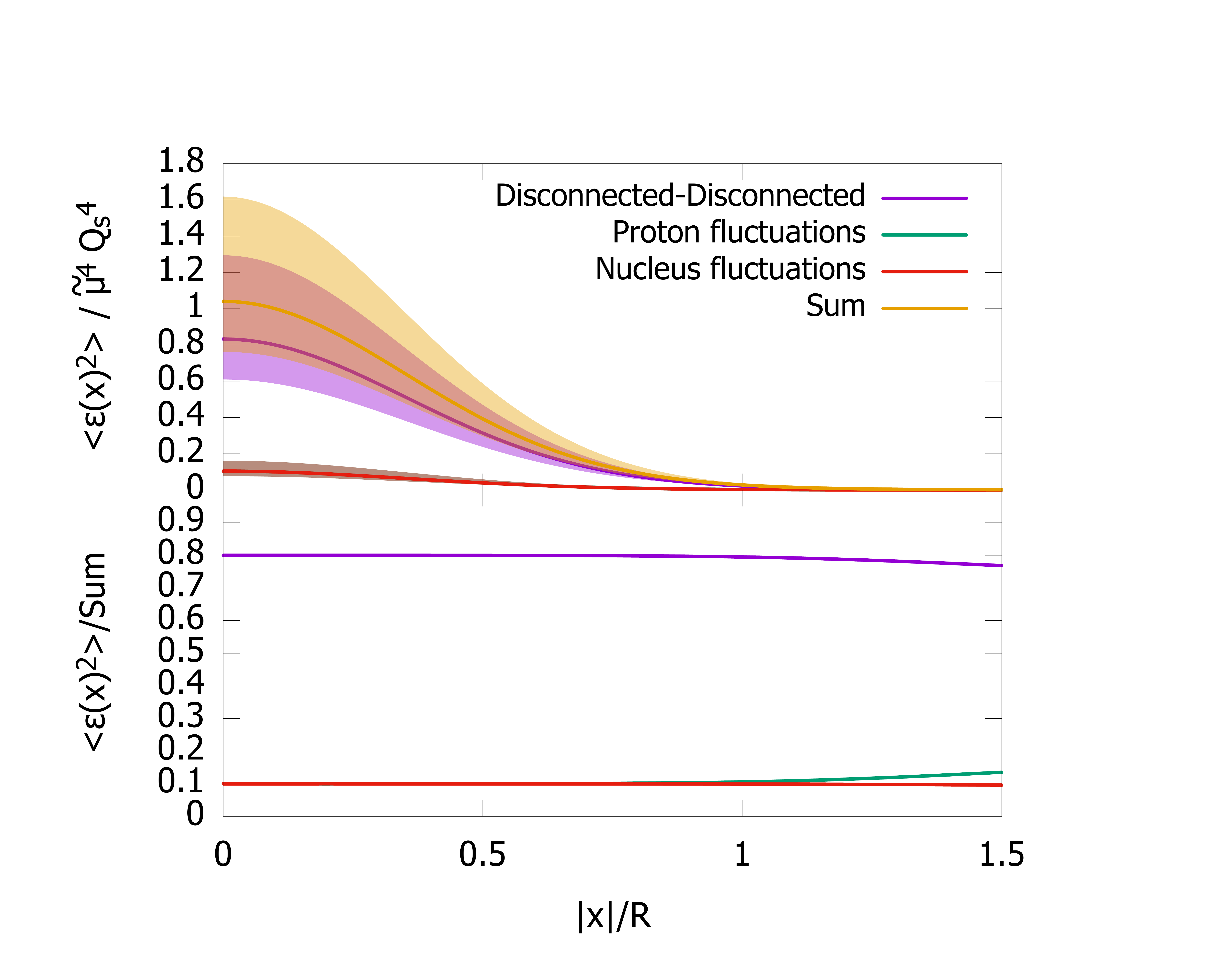}
    }
    \subfloat[\(N_q=3\)]{
    \includegraphics[width=1.1\columnwidth]{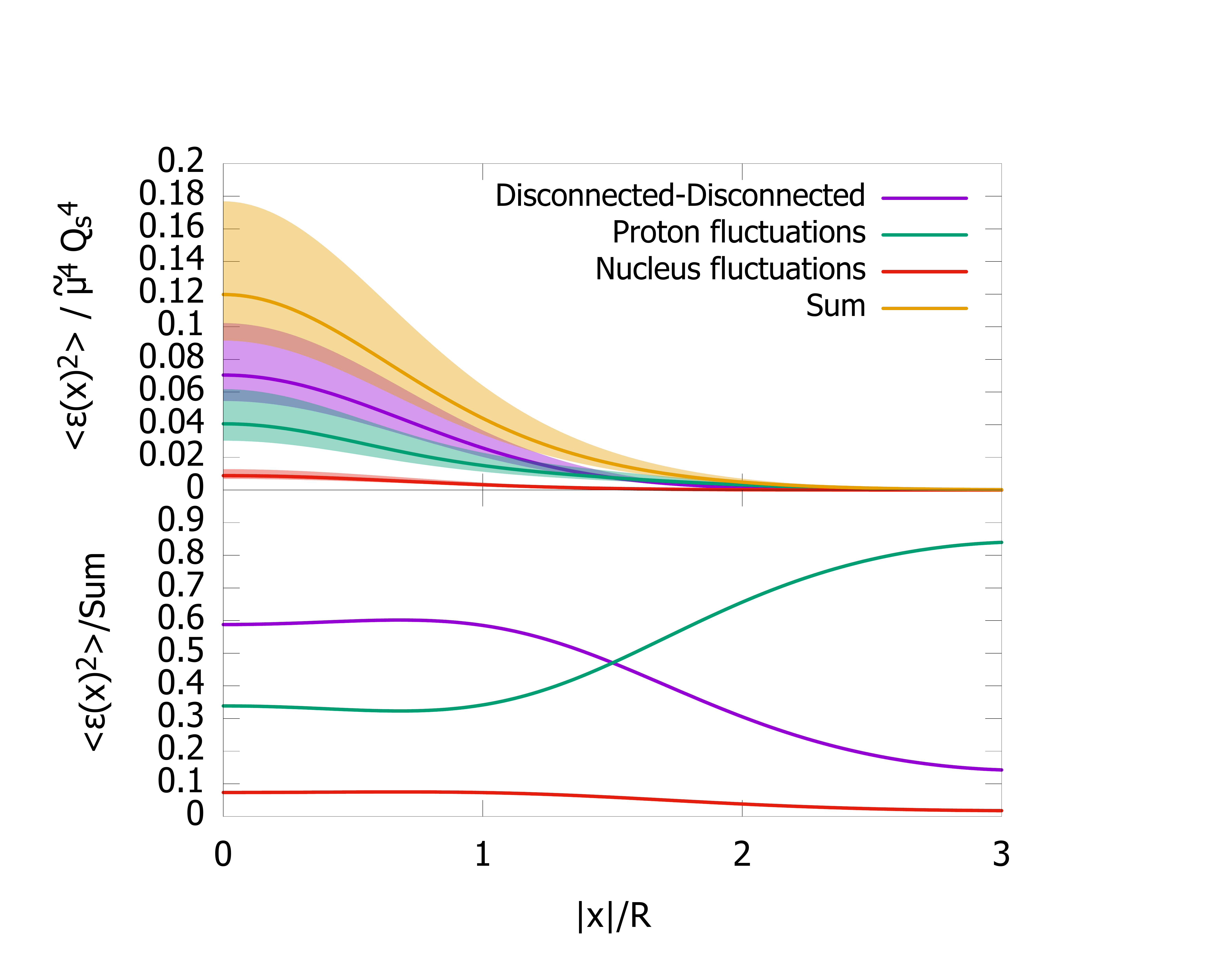}
    }
    \vspace{-3em}
    \subfloat[\(N_q=10\)]{
    \includegraphics[width=1.1\columnwidth]{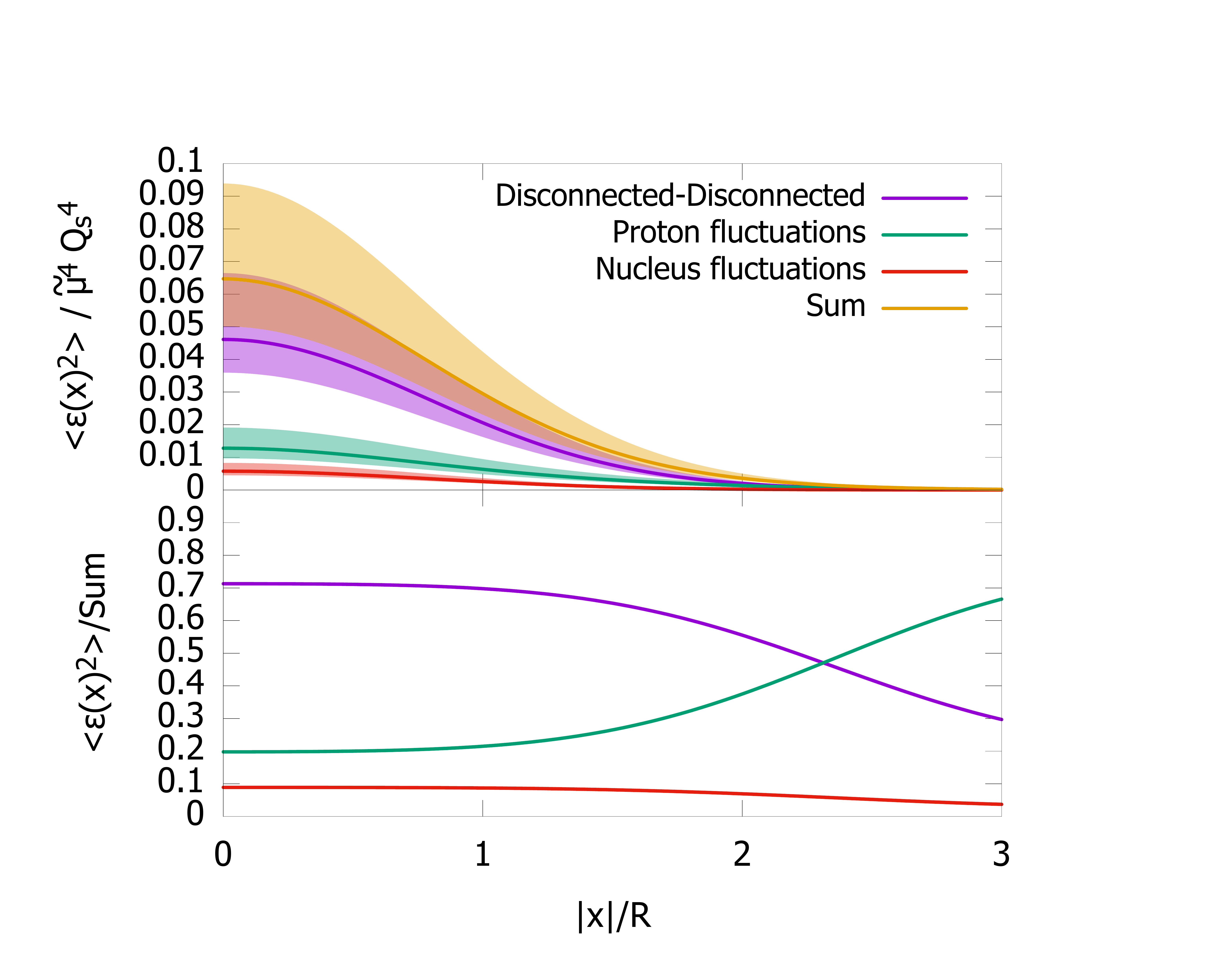}
    }
    \subfloat[\(N_q=100\)]{
    \includegraphics[width=1.1\columnwidth]{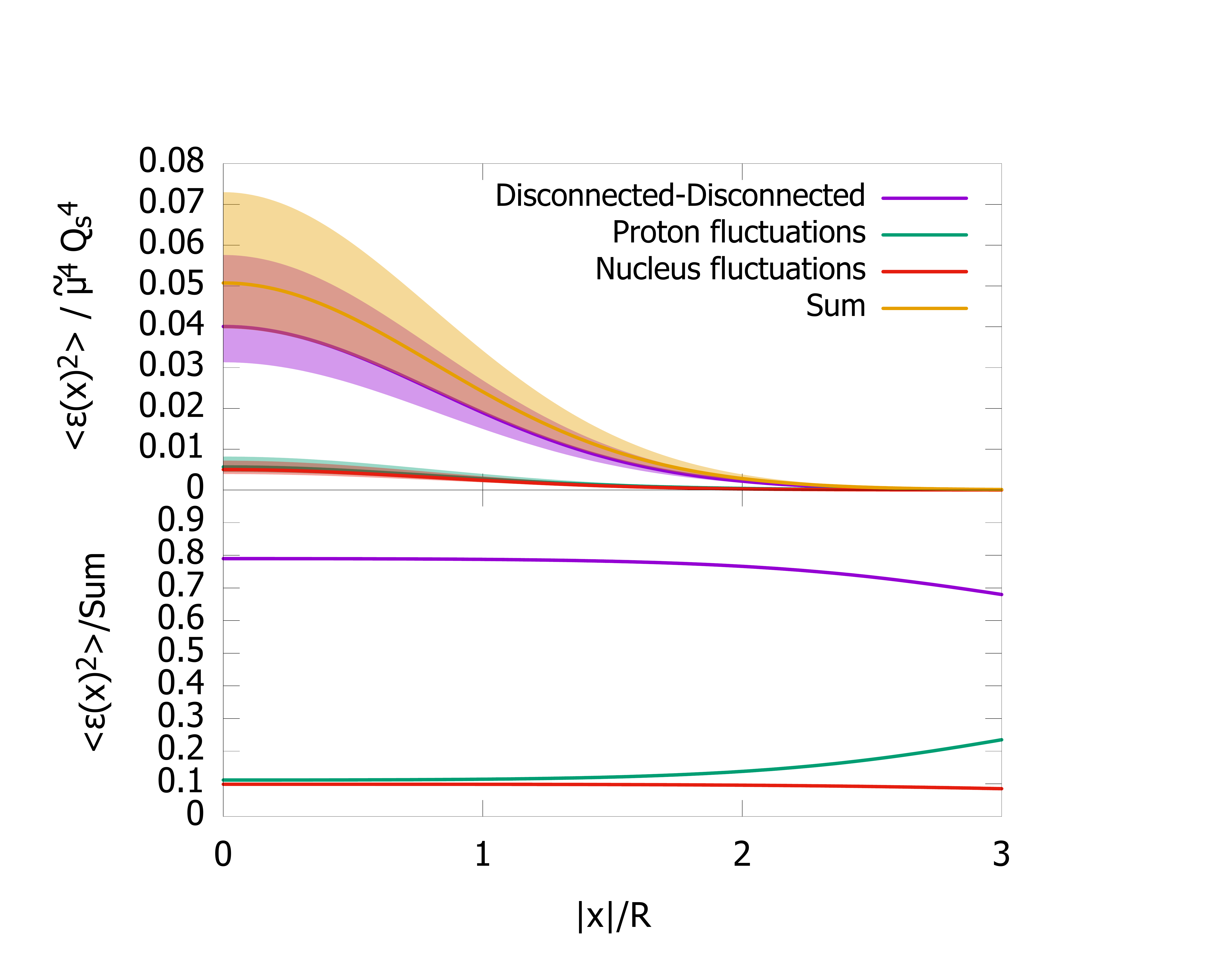}
    }
\caption{
Different contributions to the local energy density fluctuations $\langle \varepsilon(\xt)^2 \rangle$, as a function of the position $\xt$ relative to the center of mass of the proton ($\Bt=0$). The  bands correspond to varying the UV cutoff \(C_0\) by \(\pm 50\%\). The bottom plots show the relative contribution of each individual contribution.}
\label{fig:TwoPointSameC0}
\end{figure*}

\begin{figure*}[ptbh!]
    \centering
    \subfloat[\(N_q=1\)]{
    \includegraphics[width=1.1\columnwidth]{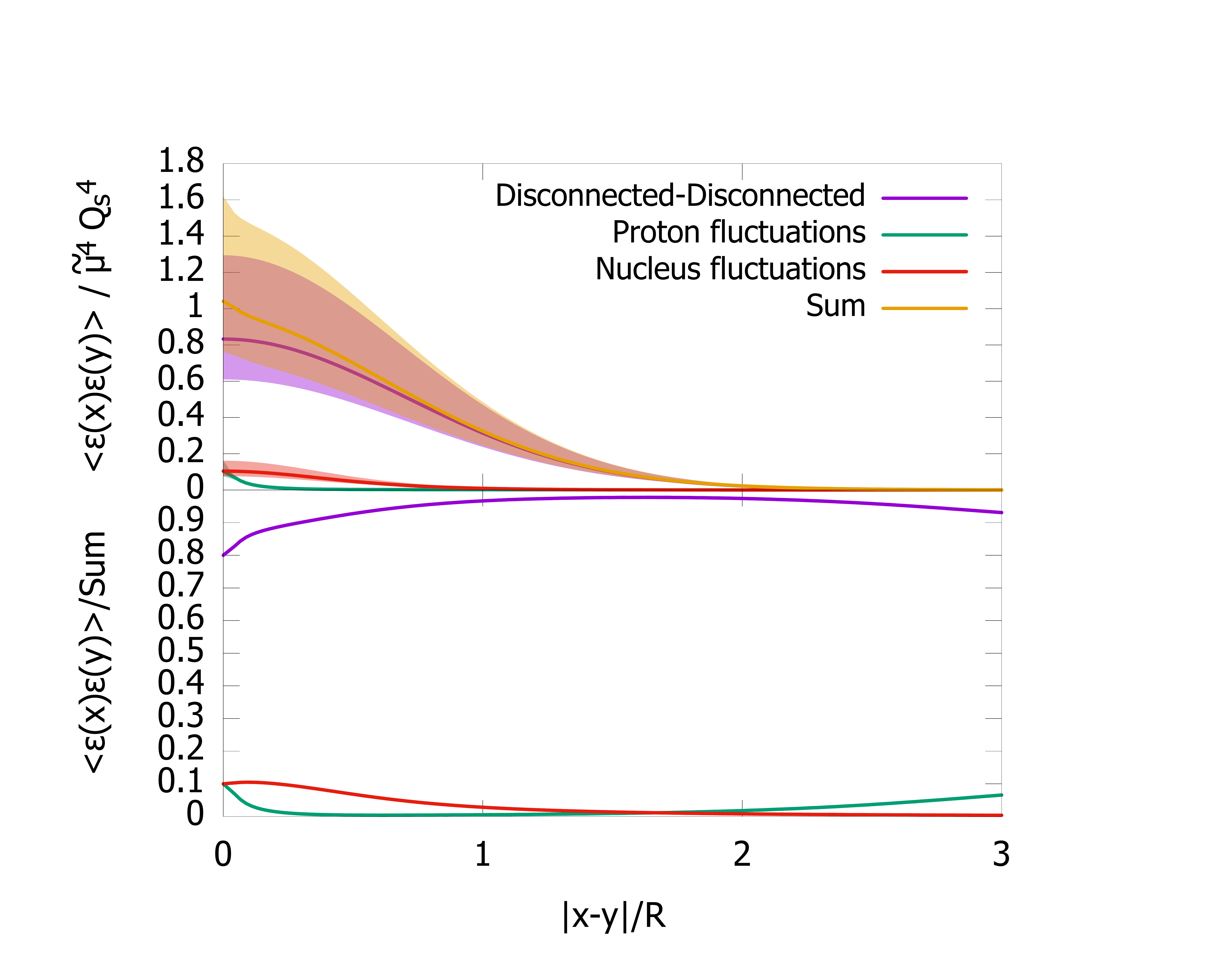}
    }
    \subfloat[\(N_q=3\)]{
    \includegraphics[width=1.1\columnwidth]{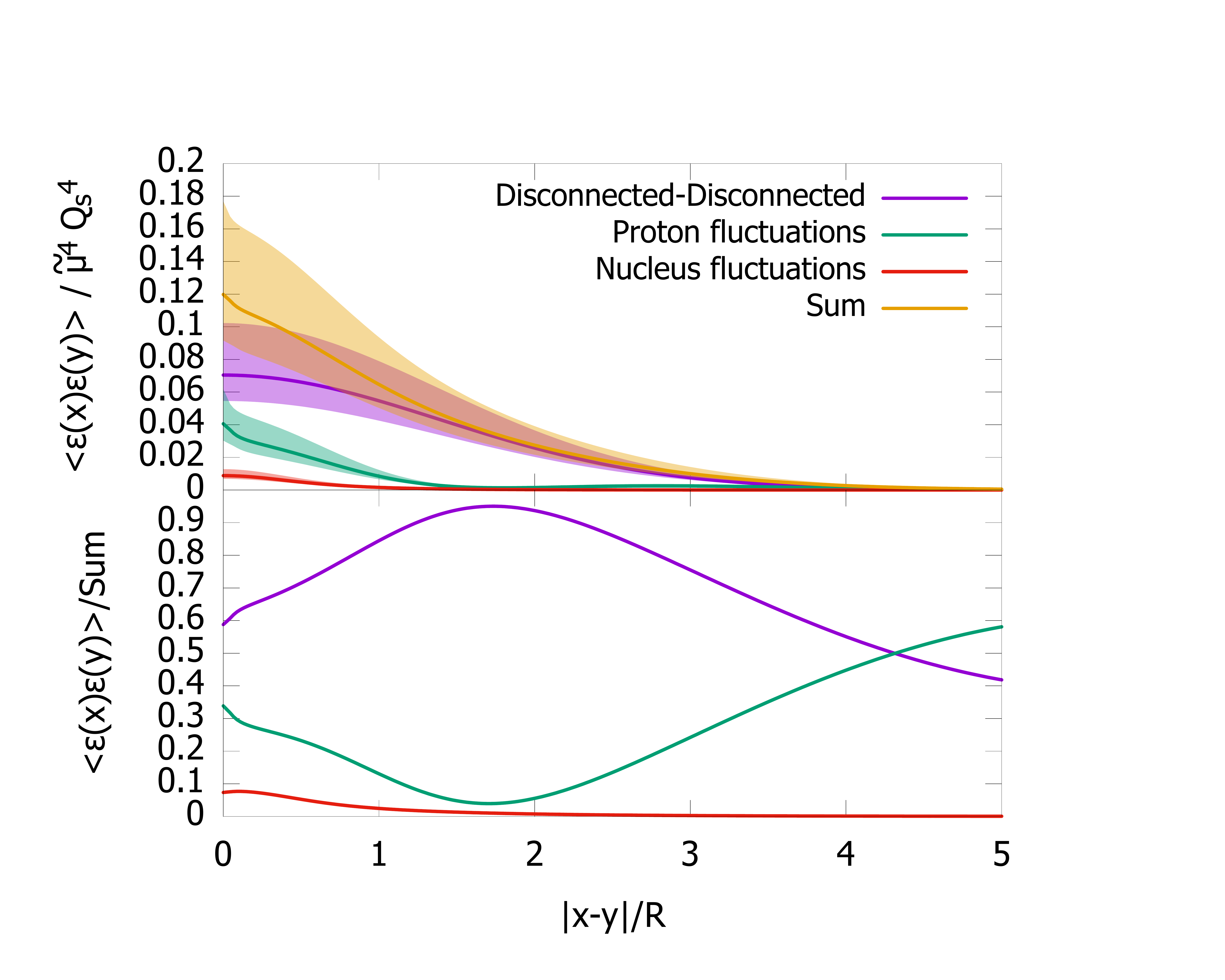}
    }
    \vspace{-3em}
    \subfloat[\(N_q=10\)]{
    \includegraphics[width=1.1\columnwidth]{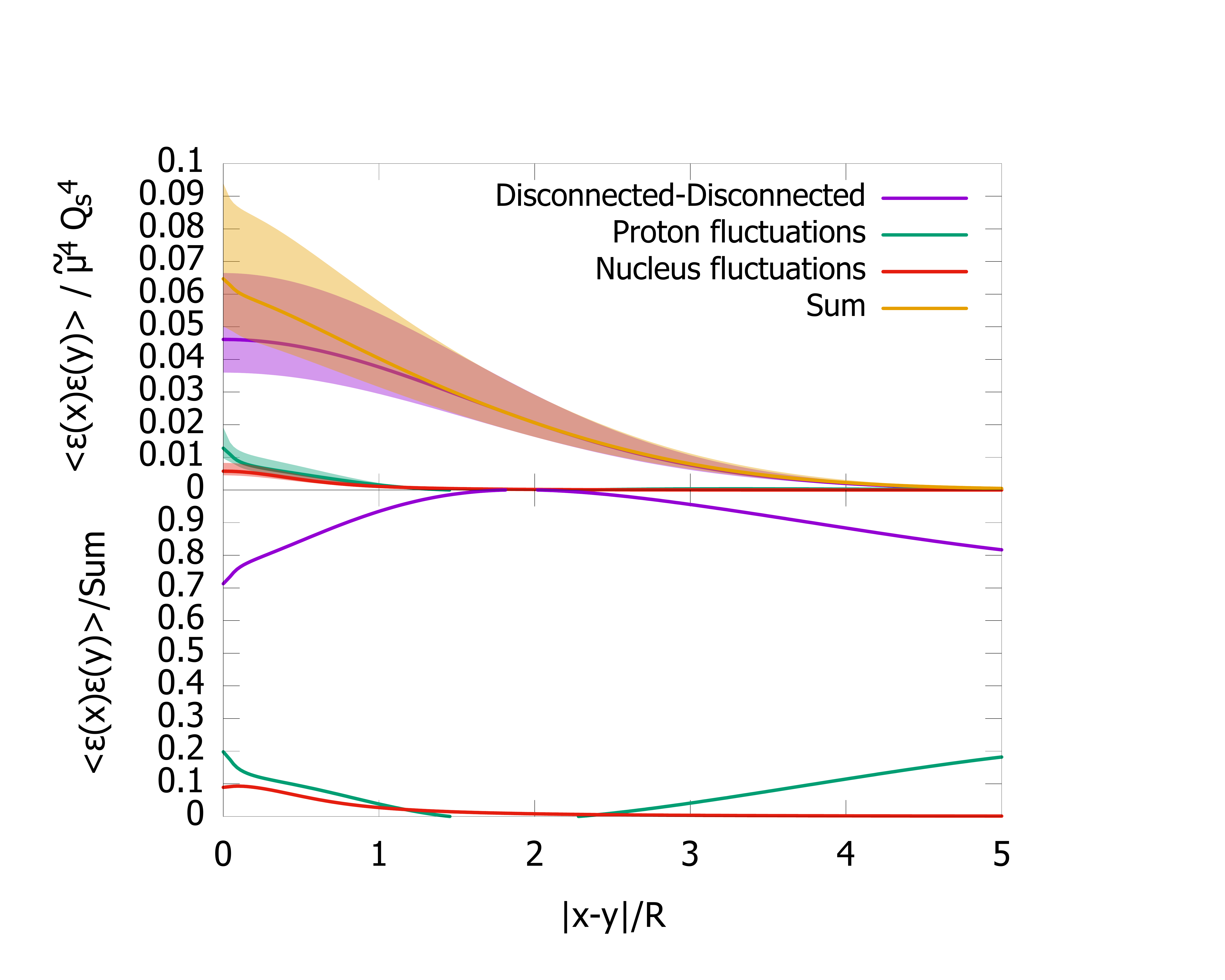}
    }
    \subfloat[\(N_q=100\)]{
    \includegraphics[width=1.1\columnwidth]{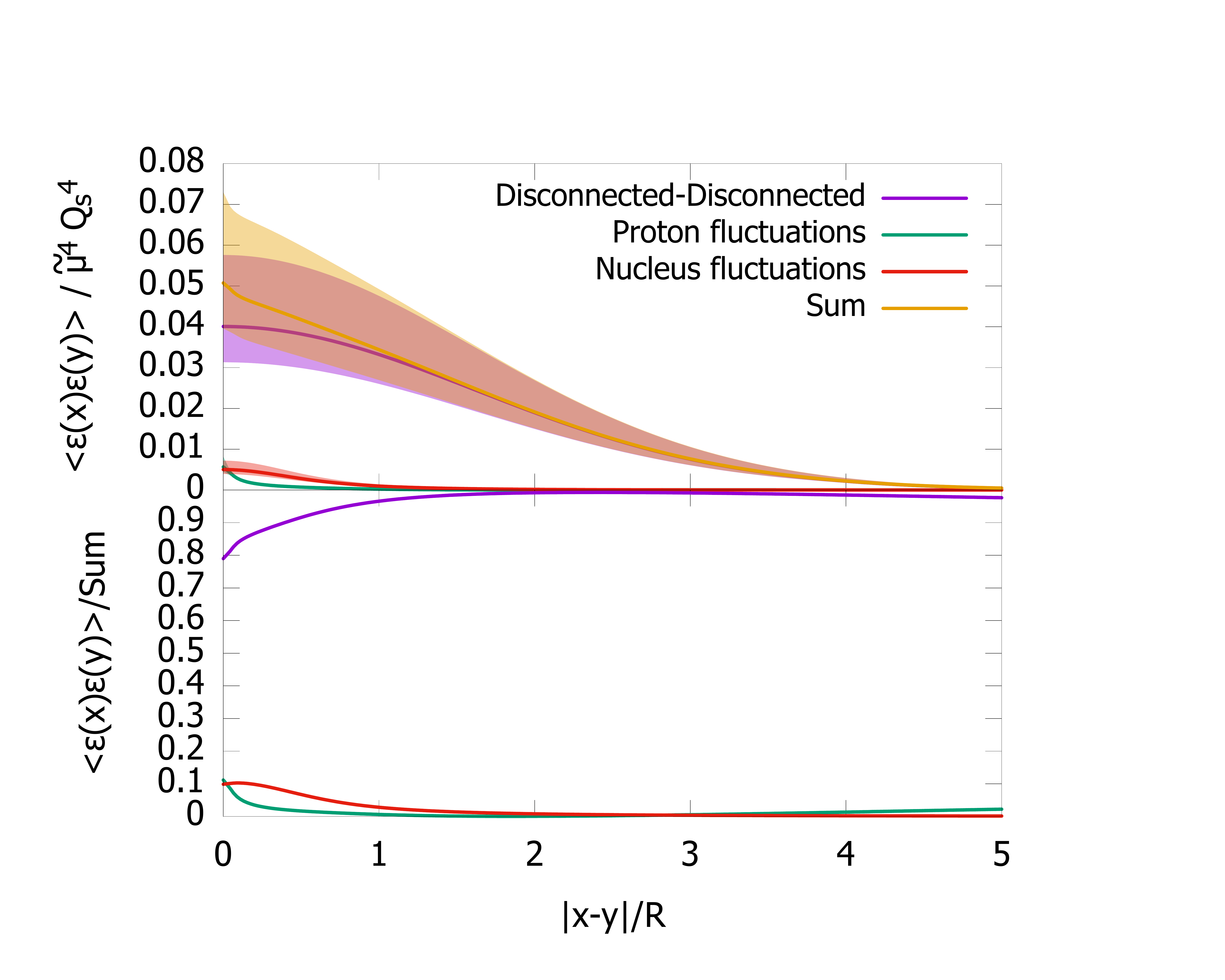}
    }
\caption{
Different contributions to the energy density correlation functions $\langle \varepsilon(\xt) \varepsilon(\yt) \rangle$, as a function of the coordinate separation $|\xt-\yt|$. The coordinates $\xt=-\yt$ are taken to be opposite to each other on a straight line through the center of the proton \((\Bt=0)\). The different panels show the results for different number of hot spots $N_{q}=1,3,10,100$. The bands correspond to variations of the UV cutoff \(C_0\) by \(\pm 50\%\). The bottom plots show  the relative contribution of each individual contribution.
}
\label{fig:TwoPointSeparateC0}
\end{figure*}

\begin{figure*}[ptbh!]
    \centering
    \subfloat[\(N_q=3\)]{
    \hspace{-2em}
    \includegraphics[width=1.15\columnwidth]{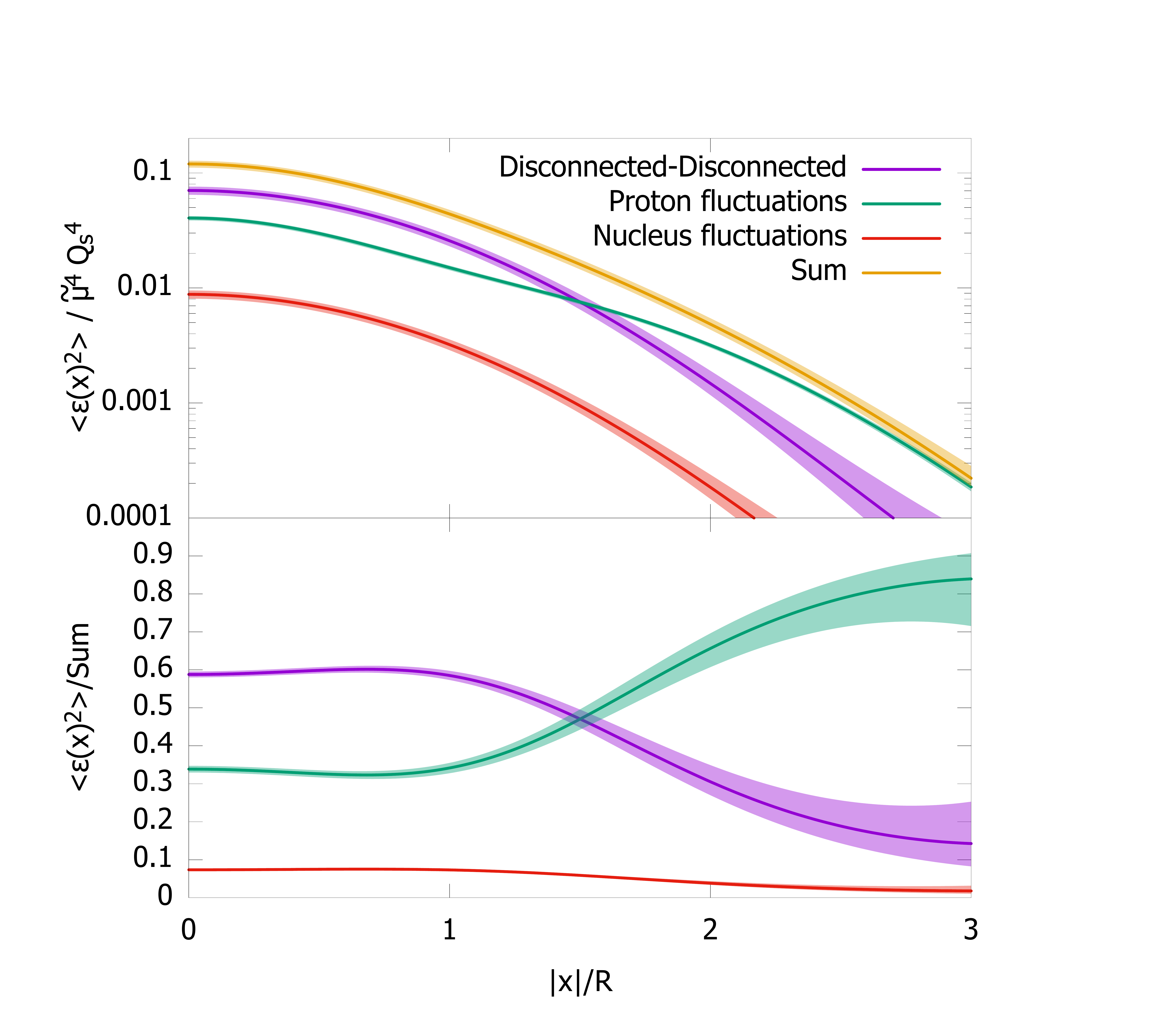}
    \hspace{-5em}
    }
    \subfloat[\(N_q=3\)]{
    \includegraphics[width=1.15\columnwidth]{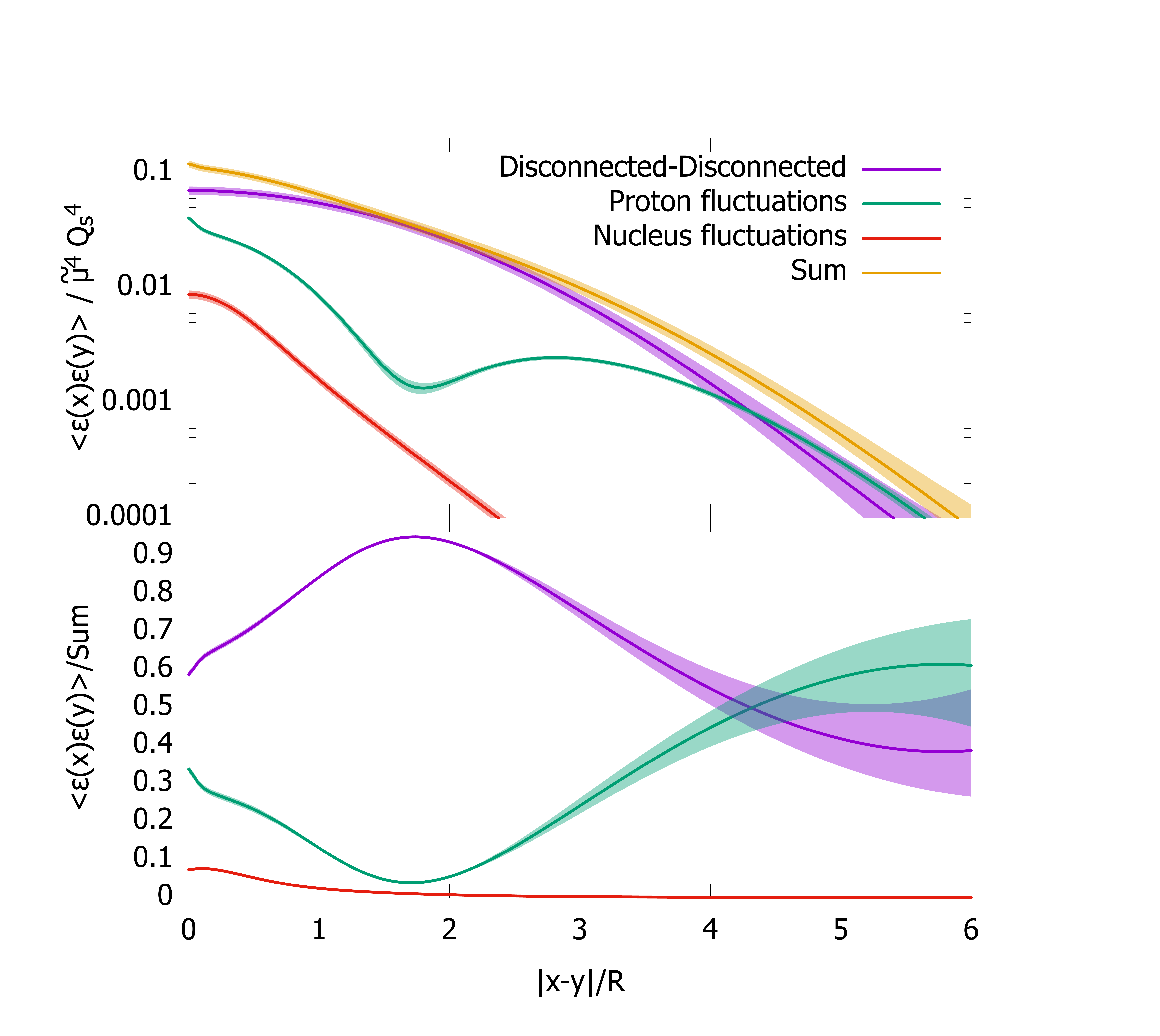}
    }
\caption{Different contributions to the local energy density fluctuations $\langle \varepsilon(\xt)^2 \rangle$ (left) and the energy density correlation functions $\langle \varepsilon(\xt) \varepsilon(\yt) \rangle$ (right), for the same configurations of the coordinates $\xt,\yt$ as in figures \ref{fig:TwoPointSameC0} and \ref{fig:TwoPointSeparateC0}. The bands are obtained by varying the IR regulator mass by \(\pm 50\%\), and the results are on a logarithmic scale to illustrate the sizable variation of the long range tails by the variation of the IR regulator mass.}
\label{fig:Nq3TwoPointMassLogPlots}
\end{figure*}

Let us now evaluate numerically the different parts of the energy density two-point function. We will plot both the energy density correlator values themselves, and below these the relative contributions of the  different parts that contribute to the total two-point function (in our approximation where we drop the connected-connected part). We will vary parameters of our model to produce an error band for the energy density correlators, with the relative contributions only plotted for the central values in this variation for clarity.  We have normalized the two-point functions in the top plots with the nucleus saturation scale \(Q_s\) and the scaled proton saturation scale parameter \(\tilde{\mu}^2\) defined as
\begin{equation}
\tilde{\mu}^2 = N_q \mu^2_0.
\end{equation}

In the rest of the paper, we use the following default parameters for plots unless stated otherwise. We always have three  colors \(\nc \equiv N =3\). The nucleus saturation scale is set to be \(Q_s=2 \text{ GeV}\), the UV cutoff is  \(C_0=0.05 \text{ GeV}^{-1}\), the IR regulating mass is \(m=0.22 \text{ GeV}\), the proton radius parameter is \(R=\sqrt{3.3} \text{ GeV}^{-1}\) and the hot spot radius is \(r=\sqrt{0.7} \text{ GeV}^{-1}\). For this work, the values for \(R\) and \(r\) were taken from \cite{Mantysaari:2016jaz}. The mass was chosen to be of the order of the QCD scale and \(C_0\) was taken to be of a scale smaller than the hot spot radius to allow for gluon fields originating from the same hot spot to overlap but still cut off the divergent behavior of the Green's functions.

We will plot the energy density correlator in two different coordinate configurations. Firstly, we set the two coordinates in the two-point function to be equal \((\xt=\yt)\) and plot the parts of the two-point function as a function of the distance from the center of the proton \((\Bt)\) divided by the proton radius \(R\). This measures the local energy density fluctuations as a function of distance. These plots are shown in figure \ref{fig:TwoPointSameC0}. Secondly we have plots of the energy density correlation, where we take a straight line through the center of the proton and let the two coordinates move, at the same rate, in opposite directions along the line. In the plots the two-point function parts are plotted as functions of the distance of these two coordinates divided by the proton radius. These plots are shown in figure \ref{fig:TwoPointSeparateC0}.

We add error bands to the plots by varying either the UV cutoff \((C_0)\) or the IR regulator \((m)\) by \(\pm 50\%\). The UV cutoff dependence seems to be much more prominent than the dependence on the IR regulator. This is not surprising, it has been long known that the energy density at exactly $\tau=0$ is logarithmically UV divergent. This divergence would, however,  disappear with the evolution to larger $\tau$~\cite{Lappi:2006hq,Lappi:2017skr}, which we do not attempt to do here\footnote{Note that, as also discussed in Refs.~\cite{Lappi:2006hq,Lappi:2017skr}, the energy density in the continuum limit of the MV-model diverges logarithmically at $\tau \to 0$, and thus cannot be developed in a power series in $\tau$. This demonstrates itself in the increasingly dramatic power law UV-divergences in the coefficients when such an expansion is nevertheless attempted~\cite{Chen:2015wia,Carrington:2020ssh}.}.  The UV-cutoff dependence is particularly clearly visible near the center of the proton, where the color charge density is largest and there is large amount of short range overlap for the gluon fields generated by these color charges. The mass, on the other hand, has more of an influence on the long range behavior on the gluon fields. The long range mass dependence is shown in figure \ref{fig:Nq3TwoPointMassLogPlots}, which indeed shows that the exact value of the mass influences where the proton fluctuations are sizable or larger than the contribution from the fully disconnected part.  We further note that the contribution from proton fluctuations to the energy density correlation, \eq\nr{eq:protonside}, is actually a sum of two contributions, a short range one originating from a single hot spot ($F_2$) and another one that is sensitive to two hot spots ($F_3$). The latter is preferentially long range, because the hot spots are typically separated by a distance $\sim R$. The interplay between these two contributions leads to an interesting structure of the correlation function with a dip and secondary maximum seen in Fig.~\ref{fig:Nq3TwoPointMassLogPlots}. These contributions are shown separately in \fig\ref{fig:ProtonTwoPointContributionsProportionalDC} in Appendix~\ref{app:numerics}.

The dependence on the number of hot spots \((N_q)\) can be seen in Figs.~\ref{fig:TwoPointSameC0} and \ref{fig:TwoPointSeparateC0}. When we only have one hot spot, in our model, it has to be in the center of the proton, and thus there are only color charge fluctuations for the proton. This results in a small contribution for both the proton and the nucleus fluctuations. With a small, larger than one, number of hot spots we start to see that, in some regions, the proton fluctuations can even become larger than the fully disconnected contribution. Especially the largest fluctuations at the edge of the proton seem to be largely due to proton shape fluctuations. In the plots where we increase the separation of the two coordinates, we can see that at small separation, the proton fluctuations are sizable in comparison to the fully disconnected contribution and at large separation they can become even larger than the disconnected contribution. However, this behavior largely disappears with a large number of hot spots. The nucleus fluctuations are purely due to the gluon field fluctuations and thus do not really vary much with the number of hot spots.
    
Overall, we see that the hot spot fluctuations are by far the dominant mechanism for fluctuations and correlations of the energy density. While the exact magnitude of their contribution depends on the values of the parameters, the numbers used here correspond quite well to the ones currently used in phenomenology. The value of the energy density depends strongly on the UV-cutoff, as expected. The dependence on the IR-regulator $m$ is less pronounced, but it is important in the long distance tail of the distribution. Consequently, it turns out to be very significant for the eccentricities, as we will see in the following Section.
    
\section{Results: eccentricities}
\label{sec:ecc}

\begin{figure}[t!]
    \centering
    \includegraphics[width=\columnwidth]{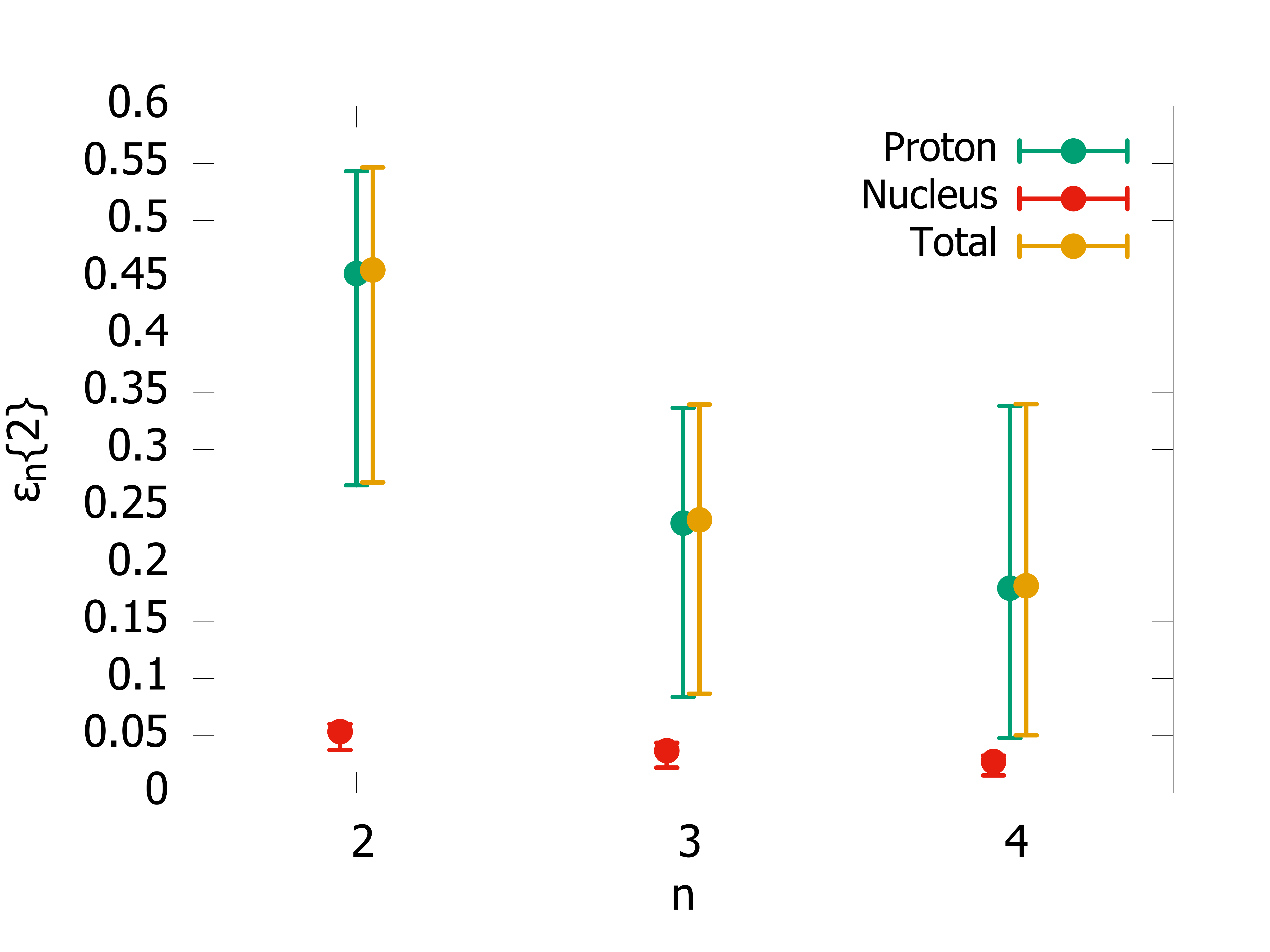}
    \caption{Eccentricities $\varepsilon_{n}\{2\}$ with \(n=2,3,4\) for a proton with three hot spots (\(N_q=3\)). The eccentricities originating from proton and nucleus fluctuations are shown separately along with the total eccentricity. The bands are obtained by varying the IR regulator mass by \(\pm 50 \%\).
    }
    \label{fig:Nq3EllipticityVsN}
\end{figure}

\begin{figure}[tbhp!]
    \centering
    \subfloat[\(n=2\)]{
    \includegraphics[width=\columnwidth]{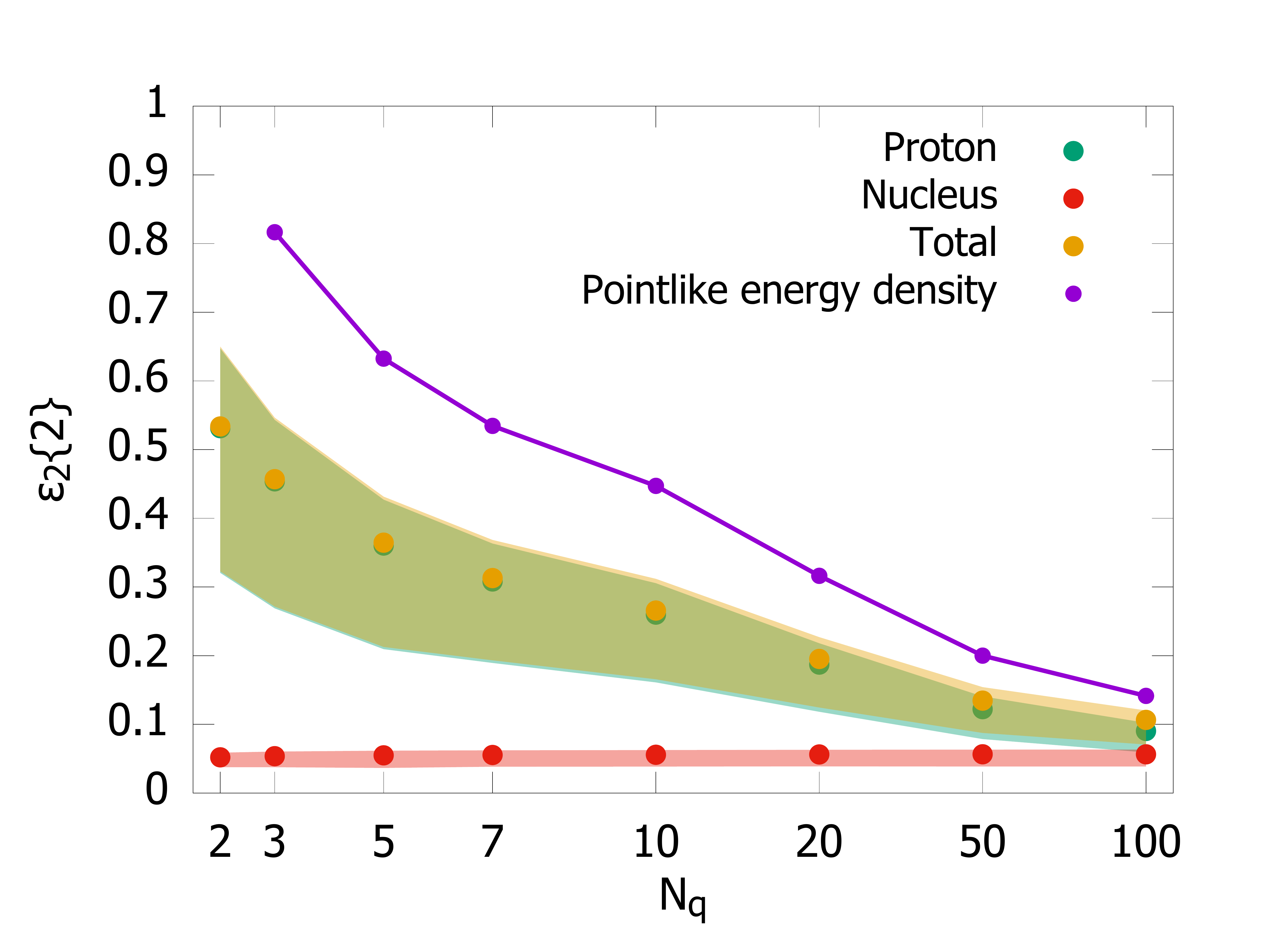}
    }
    \vspace{-2em}
    \subfloat[\(n=3\)]{
    \includegraphics[width=\columnwidth]{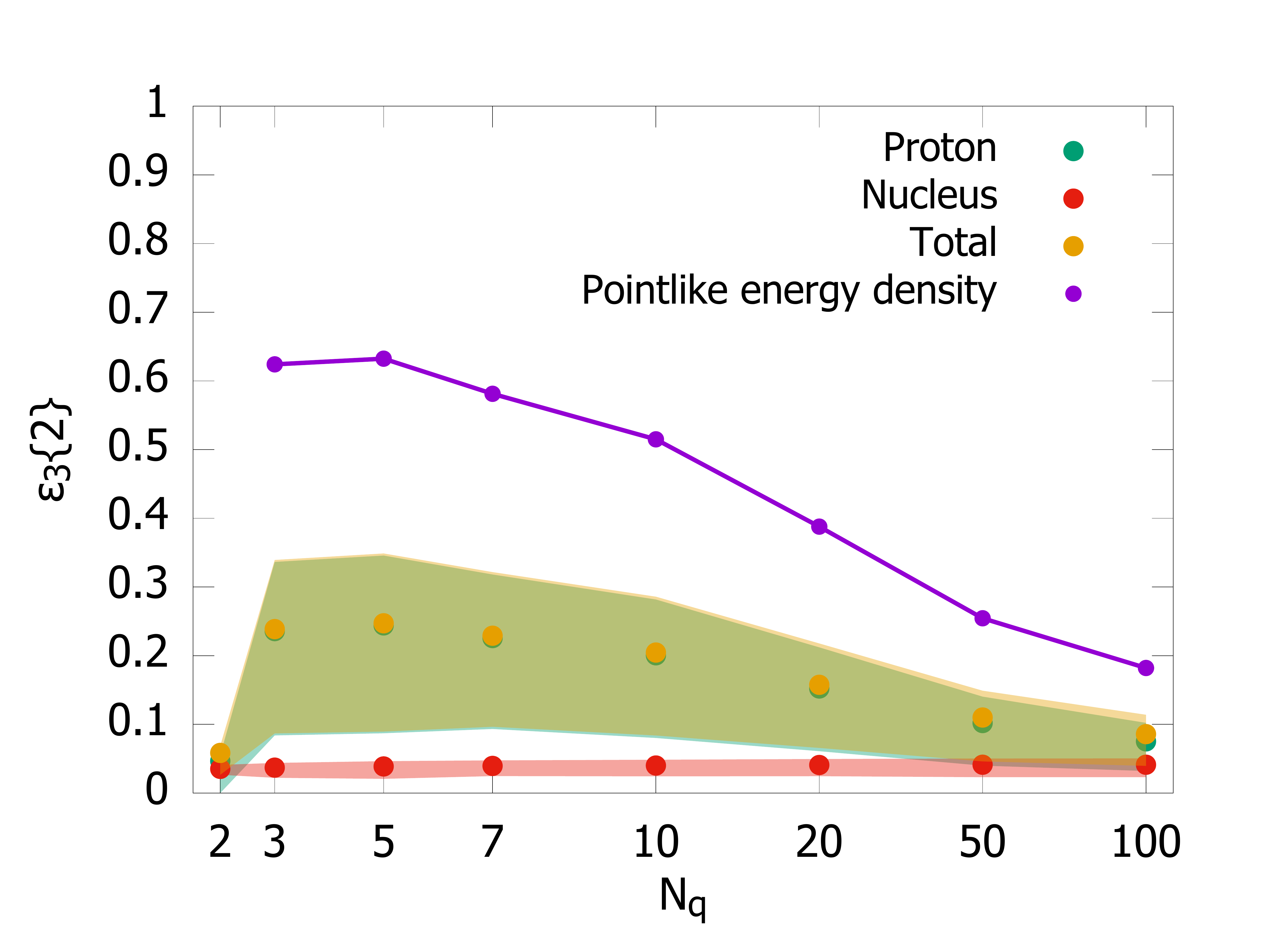}
    }
    \vspace{-2em}
    \subfloat[\(n=4\)]{
    \includegraphics[width=\columnwidth]{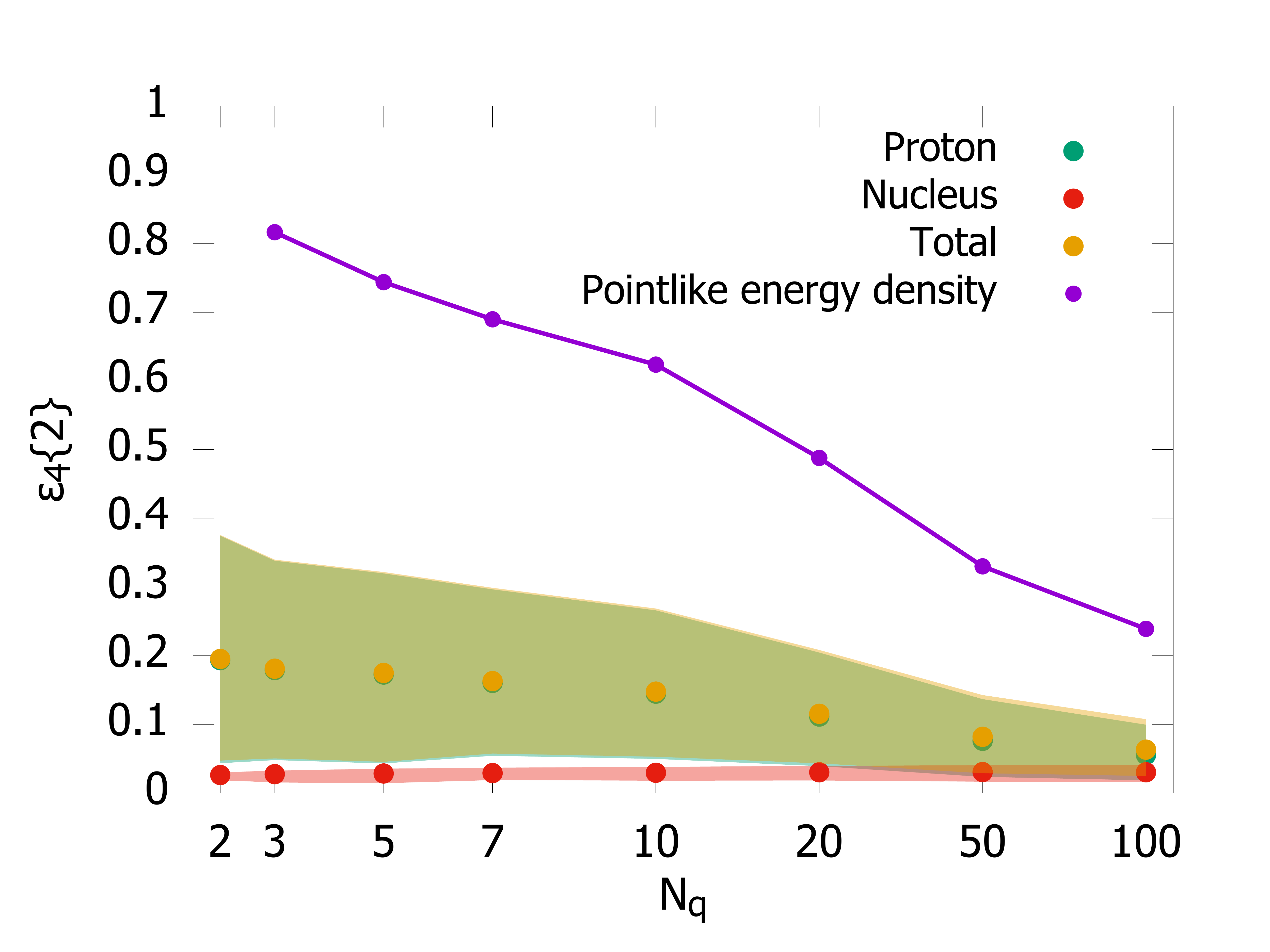}
    }
    
\caption{Eccentricities $\varepsilon_{n}\{2\}$ as functions of the number of hot spots \(N_q\) for \(n=2,3,4\) (top,middle,bottom). Confidence bands are obtained by varying the IR regulator $m$ by \(\pm 50\%\). Numerical results from the our hot spot model, are also compared to the eccentricities in a pointlike energy density model discussed in Appendix~\ref{app:delta}. }
\label{fig:n234EllipticityVsNqMass}
\end{figure}

Now let us move to computing the azimuthal eccentricities arising from the proton shape and gluon fluctuations in our model.

For a single configuration,  the deformation of the energy density field \(\varepsilon(\xt)\) from azimuthal symmetry is quantified by the dimensionless ratios known as the eccentricities, defined as \cite{Blaizot:2014nia}
\begin{equation} \label{eq:epsn}
\varepsilon _n = \frac{\int \ud^2\xt~|\xt - \Bt|^n~ e^{in\theta_{\xt-\Bt}}~\varepsilon(\xt)}{\int \ud^2\xt~ |\xt - \Bt|^n~\varepsilon(\xt)},
\end{equation}
where \(\Bt\) is the center of the event defined as
\begin{equation}
\label{eq:BEnergyDef}
\Bt = \frac{\int \ud^2\xt~\xt~\varepsilon(\xt)}{\int \ud^2\xt~\varepsilon(\xt)}.
\end{equation}
Event-by-event Monte-Carlo studies in IP-Glasma determine the center of mass $\Bt$ from the energy density as in Eq.~(\ref{eq:BEnergyDef}).  Here, in contrast, we will neglect the effect of the color charge fluctuations on the location of the center of the system, and  take the center of the collision system as the center of mass of the hot spots, as defined in \eq\nr{eq:defavg}. This is justified by the assumption that the hot spot location fluctuations are much larger than the gluon field fluctuations.
The fluctuation part of the energy density is defined for individual events as
\begin{equation} \label{eq:epsFlucDef}
\delta \varepsilon(\xt) = \varepsilon(\xt) - \langle \varepsilon(\xt) \rangle.
\end{equation}
The average of the fluctuations $\delta \varepsilon(\xt)$ vanishes. Thus to actually have sensitivity to the fluctuations in a fluctuation driven system,  we need consider a different quantity that is somehow quadratic in the energy density. One could calculate the eccentricity from the  mean square of the ratio~\eqref{eq:epsn} as
\begin{equation} \label{eq:meanSquareEccentricity}
\begin{split}
&
\varepsilon^{'}_n\{2\}^2 \equiv \langle \varepsilon _n \bar{\varepsilon}_n \rangle
\\ & =
\left<
\frac{\int \ud^2\xt \ud^2\yt |\xt - \Bt|^n|\yt - \Bt|^ne^{in(\theta _{\xt-\Bt} - \theta _{\yt-\Bt})} \varepsilon(\xt)\varepsilon(\yt)}{\int \ud^2\xt \ud^2\yt |\xt - \Bt|^n|\yt - \Bt|^n \varepsilon(\xt) \varepsilon(\yt)}
\right>
.
\end{split}
\end{equation}
Calculating the expectation value of the ratio is, however, not desirable for several reasons (see also the related discussion in Ref.~\cite{Lappi:2015vha}). Firstly it is not possible to evaluate \nr{eq:meanSquareEccentricity} based on just the two point function of the energy density that we have available in our model. Nevertheless we consider the energy density correlator as the more fundamental quantity that should contain the relevant physics. We could only evaluate \nr{eq:meanSquareEccentricity} in the regime of small fluctuations, but this might not be a valid approximation for us due to the large fluctuations arising from the hot spot model.   More generally in the CGC formalism, or in quantum mechanics in general, it is natural to calculate expectation values of operators that are positive integer powers of physical observables, such as the energy density. Also in experimental determinations of flow coefficients $v_n$, it is not immediately obvious to us whether one is closer to calculating ratios of expectation values or expectation values of ratios; this depends on how the event-by-event varying total multiplicity is treated in the analysis. In light of this discussion we will here define the eccentricity through the more natural quantity, namely the one obtained from the ratio of the averages of azimuthal harmonics of the energy density two point function. The definition of the eccentricity that we use is
\begin{equation} \label{eq:EccTwoPointDenom}
\begin{split}
&
\varepsilon_n\{2\}^2
\\ & =
\frac{\int \ud^2\xt \ud^2\yt |\xt-\Bt|^n|\yt-\Bt|^n e^{in(\theta _{\xt-\Bt} - \theta _{\yt-\Bt})}
\langle
\varepsilon(\xt)\varepsilon(\yt)
\rangle}
{\int \ud^2\xt \ud^2\yt |\xt-\Bt|^n|\yt-\Bt|^n
\langle
\varepsilon(\xt) \varepsilon(\yt)
\rangle}
.
\end{split}
\end{equation}
Due to the infinitely large homogeneous nucleus in our model, this quantity is actually independent of \(\Bt\), and we can set \(\Bt=0\) without loss of generality.

\begin{figure}[tbp!]
    \centering
    \includegraphics[width=\columnwidth]{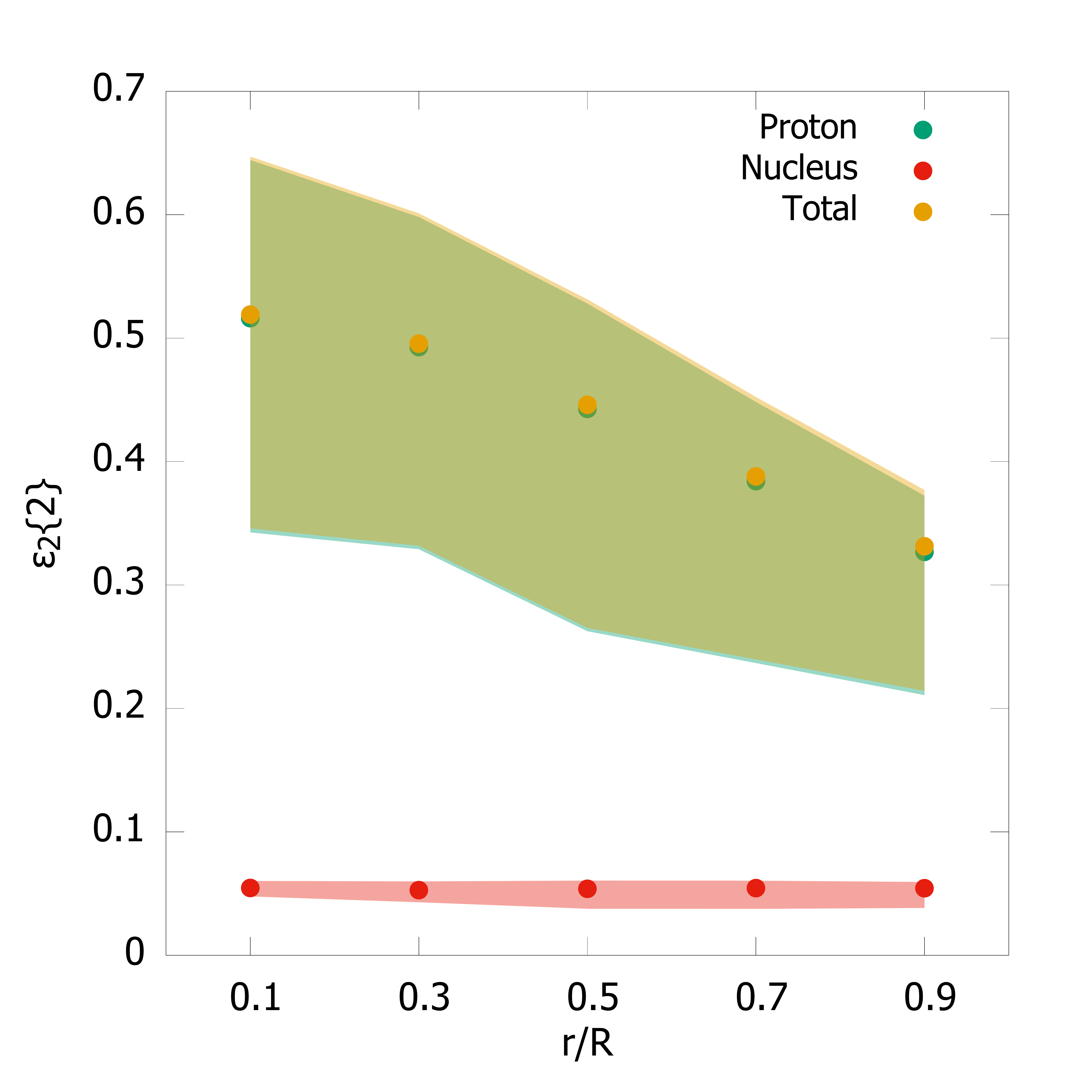}
    \caption{Elliptic eccentricity $\varepsilon_{2}\{2\}$ for a proton with $N_{q}=3$ hot spots, as a function of the hot spot radius \(r\), keeping the proton size \(R\) fixed. Confidence bands are obtained by varying the IR regulator mass by \(\pm 50\%\).}
    \label{fig:n2Nq3rOverR}
\end{figure}

We note that our definition in Eq.~\nr{eq:EccTwoPointDenom} is equivalent to the expectation value of the ratio \eqref{eq:meanSquareEccentricity} when the fluctuations of energy density \eqref{eq:epsFlucDef} are small. In this case one can expand the  eccentricities to the first order in the fluctuations, which leads to the approximation used e.g. in Ref.~\cite{Blaizot:2014nia}
\begin{equation} \label{eq:EccApprox}
\begin{split}
&
\varepsilon^{'}_n\{2\}^2_{\text{Approx}}
\\ & =
\frac{\int \ud^2\xt \ud^2\yt |\xt-\Bt|^n|\yt-\Bt|^ne^{in(\theta _{\xt-\Bt} - \theta _{\yt-\Bt})}
\langle
\varepsilon(\xt)\varepsilon(\yt)
\rangle}
{\int \ud^2\xt \ud^2\yt |\xt-\Bt|^n|\yt-\Bt|^n
\langle
\varepsilon(\xt)
\rangle
\langle
\varepsilon(\yt)
\rangle}
,
\end{split}
\end{equation}
where the difference with respect to \eq\nr{eq:EccTwoPointDenom} is that the two-point function in the denominator gets replaced by the product of one-point functions. This approximation has been used in \cite{Giacalone:2019kgg} in the case of A-A collisions when considering only the color fluctuations of the nuclei. Now that we incorporate the shape fluctuations to the proton through the hot spot model resulting in larger fluctuations, we can not  a priori be sure that the use of this small fluctuation approximation is valid in our case. However, in practice the approximation~\nr{eq:EccApprox} is quite close to our full result, as demonstrated in Appendix~\ref{app:numerics}.

Our main results for the eccentricities $\varepsilon_{2}\{2\}$,~$\varepsilon_{3}\{2\}$ and $\varepsilon_{4}\{2\}$ are shown in \figs\ref{fig:Nq3EllipticityVsN}, \ref{fig:n234EllipticityVsNqMass} and~\ref{fig:n2Nq3rOverR}. We compute the contribution to the eccentricities separately for proton and nucleus side fluctuations, noting that the sum of the proton and nucleus side fluctuation eccentricities does not equal to the total eccentricity, since the contributions are added quadratically  (c.f. Eq.~(\ref{eq:EccTwoPointDenom})). 

Figure~\ref{fig:Nq3EllipticityVsN} shows the eccentricities for $n=2,3,4$ for the default values of our parameters with three hot spots ($N_q=3$). We see that the proton fluctuation contribution completely dominates the eccentricities. The error bars, resulting from a variation of the parameter $m$ by $\pm 50\%$, show that the result is very strongly dependent on this infrared cutoff, especially for the higher harmonics.
It is important to to emphasize that, in contrast to the impression one would get from the two-point energy density plots, the dependence on the UV and IR regulators has changed. The UV regulator mainly affects the normalization of the energy density correlator, but has very little influence on its coordinate dependence. Thus it has very little effect on the eccentricities, as quantified explicitly in \fig\ref{fig:n234EllipticityVsNqC0} in Appendix~\ref{app:numerics}. Conversely the IR regulator affects the long range physics to which the eccentricities are sensitive due to the large $\sim|\xt-\Bt|^{n}$ weight of the long range tails, resulting in the sizeable uncertainties of the eccentricities seen in Fig.~\ref{fig:Nq3EllipticityVsN}.

We then focus on the  dependence of the eccentricities on the number of hot spots, which is plotted in \fig\ref{fig:n234EllipticityVsNqMass}. One can see that the overall trend for the proton fluctuation part of the eccentricity is such that it decreases as the number of hot spots increases. This can be explained by the reduction of event-by-event fluctuation due to a large amount of independent hot spots, which makes the events smoother and more azimuthally symmetric. We also see that the nucleus fluctuations are not as sensitive to the number of hot spots, as expected.  The proton fluctuations remain the dominant source of eccentricity up to very high numbers of hot spots, and the nucleus fluctuation contribution is negligible. An important exception to this systematic trend is $\varepsilon_3$ at $N_q =2$, where the hot spots are forced to be exactly back-to-back, and the contribution to $\varepsilon_3$ is  solely due to color charge fluctuations.

Next we study the effect of the size $r$ of an individual hot spot. Figure~\ref{fig:n2Nq3rOverR} shows the second order (\(n=2\)) eccentricity as a function of the hot spot size \(r\) divided by the proton size parameter \(R\) (obtained keeping \(R\) constant and varying \(r\)). Firstly we notice that the dependence on the mass is still rather large in comparison to the exact value of the hot spot radius. The hot spot radius does have a sizable effect on the proton fluctuation part, with larger hot spots making the system smoother and consequently less eccentric, as expected.

Overall, we find that the values of the eccentricities can be substantial, with $\varepsilon_2\{2\}$ reaching above 0.5 for a small number of hot spots, with the exact value very strongly depending on $m$. In order to set these values in perspective, we have also constructed a model of an energy density constituted by $N_q$ exactly pointlike hot spots, correlated only by the requirement that their center of mass is at a fixed coordinate.  This model, inspired by but not equivalent to the work in Refs.~\cite{Blaizot:2014nia,Giacalone:2019kgg}, would correspond to the limit $r\to 0,m \to \infty$ of our hot spot model, where energy density consists of a superposition of delta function like peaks. In this limit, the eccentricities can in fact be calculated analytically, and the results are provided in Appendix~\ref{app:delta}. Figure~\ref{fig:n234EllipticityVsNqMass} also shows a comparison of the eccentricities to this pointlike energy density model. One sees that the eccentricities of the hot spot model are still a factor of $\sim 2$ smaller. We therefore conclude that the smoothing provided by a finite hot spot size $r$ and the Coulomb tails of the color fields regulated by $m$, 
still has a significant impact on the eccentricities, and a realistic color field calculation cannot be accurately described by an energy density consisting of just a finite number of delta function peaks.

\section{Conclusions}
\label{sec:conc}

We have constructed an analytical model for calculating the initial energy density and its fluctuations in the initial stages of a high energy hadronic collision. Our model is formulated in an extremely asymmetric dilute-dense limit, which enables an analytical calculation  of the energy density correlator. In this limit, we can very explicitly see the influence of different physical aspects of the model (a philosophy very strongly advocated recently e.g. in Ref.~\cite{Nagle:2018ybc}). The most important aspect that we have focused on here is the role of hot spot fluctuations in the dilute proton projectile, which have recently been understood to be crucial for understanding the geometry of the initial stages of high energy collisions~\cite{Mantysaari:2020axf}. We have also quantified  the extremely important role for the eccentricities played by the infrared regulator parameter $m$. In contrast, the UV divergence present in the glasma fields in the MV model at $\tau=0$ has very little influence on the eccentricities. 

In our model, with parameters taken to have what we believe as reasonable values for a central proton-nucleus collision, the proton hot spot fluctuations are much more important for the eccentricities than the color field fluctuations in the nucleus. In addition to the IR regulator for the Coulomb tails $m$, their effect strongly depends on the hot spot size $r$, and the number of hot spots $N_q$, which we treat as a free parameter. In order to enable a more realistic comparison to experimental data in the future, it would be important to include geometric (nucleon position and internal nucleon structure) fluctuations also in the target nucleus. We have not done so here, to be able to more cleanly demonstrate the effect of the separate sources of fluctuations. Another  future development is the use diffractive $ep$ scattering data to constrain the parameters of our model. For example, we believe there is a significant degeneracy in the numerical values of the hot spot size $r$ and the IR cutoff $m$. Our model  for the proton color fields is simultaneously versatile enough to describe the relevant physics, but simple enough to be analytically solvable.  We believe that this combination will make it useful for a systematic study of such effects, which we plan to pursue in a future publication.

Another future avenue of study wold be to combine this initial condition with an analytical approach for the $\tau$ dependence, e.g. through some suitable resummation of an approach like the one of Refs.~\cite{Chen:2015wia,Carrington:2020ssh}. It would also be interesting to extend our calculation to 4th order correlators of the energy density. This would be possible in principle in our framework, but especially on the nucleus side the calculations would become rather complicated as the required 16-point fundamental Wilson line correlators have never been calculated so far.

\textit{Acknowledgements:} We thank A. Soto Ontoso, B. Schenke, H. Mäntysaari and P. Guerrero Rodríguez for insightful discussions. We gratefully acknowledge support by the Deutsche Forschungsgemeinschaft (DFG) through the grant CRC-TR 211 “Strong-interaction matter under extreme conditions” Project number 315477589 (S.S.) and  the Academy of Finland, project No. 321840 (TL). This work is supported  by the European Research Council, grant ERC-2015-CoG-681707.  The content of this article does not reflect the official opinion of the European Union and responsibility for the information and views expressed therein lies entirely with the authors. 

\appendix

\section{Computation of the nucleus side Wilson line correlators}
\label{app:gaussian}

We wish to evaluate the nucleus side Wilson line correlators in a Gaussian model for an infinite homogenous nucleus. The Wilson lines can be written as
\begin{equation}
U(\xt)=\mathcal{P}_+ \exp \Big [ -ig \int \ud z^+ \ud^2\zt G(\xt - \zt) \rho _a(z^+,\zt)t^a \Big ]
\end{equation}
and are defined through the series expansion of the exponential. In this Appendix we follow the method presented in \cite{Blaizot:2004wv} where it was used to compute 4-point Wilson line correlators.

As  a first illustration of the method let us discuss the simpler case of computing the 2-\(\alpha\) correlator needed for the energy density and the nucleus disconnected part. First we have to choose a basis for the computation. The exact choice depends on the correlator we are computing. In our case we can choose the basis to be built from basic building blocks consisting of two Wilson lines with open fundamental indices
\begin{equation} \label{eq:CorrelatorBuildingBlock}
[U_{\xt_1}U^{\dagger}_{\xt_2}]_{ij}.
\end{equation}

Now the actual correlator we wish to evaluate is 
\begin{equation}
\langle [U_{\xt_1}U^{\dagger}_{\xt_2}]_{k_1 k_2} [U_{\xt_3}U^{\dagger}_{\xt_4}]_{k_3 k_4} \rangle.
\end{equation}
The main idea in the computation is to expand the Wilson line exponentials into series and then taking all the possible contractions of two color charge densities \(\rho\), as permitted by the Gaussian model for the CGC weighting functional. The main ingredient for the calculation, the 2-point function of \(\rho\), reads
\begin{equation}
\langle \rho _a (x^+,\xt) \rho _b (y^+,\yt) \rangle  = \delta ^{ab} \delta(x^+ - y^+) \delta(\xt - \yt) \mu^2(x^+),
\end{equation}
where \(\mu^2(x^+)\) is the density of color charges. One also has to take into account the path-ordering of the Wilson lines, which results in all but two types of contractions to give a zero. The tadpole type contractions factorize out, and can be recombined to the final result later. On the other hand, the contractions that connect two different Wilson lines result in state transitions and are the part that is nontrivial to compute. Organizing the calculation in such a way that we start to contract the Wilson lines starting from LC time \(z^+=\infty\) and progressing towards \(z^+=-\infty\), the state transitions do not care about how the Wilson lines have been contracted in the distant past. In fundamental representation contractions, which we are right now dealing with, the color algebraic structure of the states changes according to the well known Fierz identity
\begin{equation}
t^a_{ij}t^a_{kl} = \frac{1}{2}\delta^{il}\delta^{jk} - \frac{1}{2N}\delta^{ij}\delta^{kl}.
\end{equation}
Considering this, we can see that the states that mix with our correlator are the ones where we cut open the \(z^+=\infty\) side of the Wilson line "dipoles" and connect them in every possible way while still connecting only one Wilson line to one daggered Wilson line. Thus with 4 Wilson lines we have two possible basis states which can be chosen to be
\begin{equation}
\begin{split}
\mathcal{W}_1 & = [U_{\xt_1}U^{\dagger}_{\xt_2}]_{k_1 k_2} [U_{\xt_3}U^{\dagger}_{\xt_4}]_{k_3 k_4}
\\
\mathcal{W}_2 & = [U_{\xt_1}U^{\dagger}_{\xt_4}]_{k_1 k_4} [U_{\xt_3}U^{\dagger}_{\xt_2}]_{k_3 k_2}.
\end{split}
\end{equation}

Let us now express a general correlator in this basis as a vector, where standard basis vectors represent the different basis states in such a way that \(\mathcal{W}_i \rightarrow e_i\). In this way, the general correlator can be written as
\begin{equation}
A \times \mathcal{W}_1 + B \times \mathcal{W}_2 
\rightarrow
\begin{bmatrix}
A \\
B
\end{bmatrix}.
\end{equation}

Next one needs to compute the transition matrix that describes how the basis states evolve as one contracts two \(\rho\)s, each from a different Wilson line, in every possible way. This matrix gains more and more powers as we contract more and more \(\rho\)s. These powers of the matrix exponentiate and one gets an expression where the last nontrivial task is to compute the matrix exponential. The expression still requires initial conditions for the basis states, which we again express as a vector. This means that we need to evaluate the basis states when all the \(\rho\)s have been contracted i.e. when the Wilson lines go to identity. In our case, they tell how the fundamental indices of the Wilson lines are connected in the end. For example the initial condition vector for our chosen basis states would be
\begin{equation}
\begin{bmatrix}
\delta ^{k_1 k_2} \delta ^{k_3 k_4} \\
\delta ^{k_1 k_4} \delta ^{k_2 k_3}
\end{bmatrix}
\end{equation}
Now the full expression for the 4-point correlator we need for the nucleus can be written as
\begin{equation}
\label{eq:corr2x2}
\begin{split}
&
\langle [U_{\xt_1}U^{\dagger}_{\xt_2}]_{k_1 k_2} [U_{\xt_3}U^{\dagger}_{\xt_4}]_{k_3 k_4} \rangle
\\ & 
=
\underbrace{
\begin{bmatrix}
 \delta^{k_1 k_2}\delta^{k_3 k_4} \\
 \delta^{k_1 k_4}\delta^{k_2 k_3} 
\end{bmatrix}
^{\mathrm{T}}
}_{\text{The initial conditions}}
\underbrace{
e^{M_{2\times2}}
}_{\text{The transitions}}
\underbrace{
\begin{bmatrix}
1  \\
0
\end{bmatrix}
}_{\text{The correlator}},
\end{split}
\end{equation}
where the matrix $ M_{2\times2}$ is proportional to the density that has been integrated over the longitudinal coordinate
\begin{equation}
\mu ^2 \equiv \int _{-\infty} ^{\infty} dz^+ \mu ^2 (z^+).
\end{equation}
The expression \nr{eq:corr2x2} for the correlator can be broken into 3 parts: the initial conditions and the transition matrix, which are universal for any correlator that can be expressed in our chosen basis, and the correlator vector, which depends on the actual correlator we wish to compute.

Now that the expression we are dealing with has fundamental representation \(SU(N)\) group generators contracting the correlators, we can just contract both sides of this equation with the \(t^{a}_{k_2 k_1} t^{a'}_{k_4 k_3}\). Doing this we get
\begin{equation}
\langle \text{Tr}[t^aU_{\xt_1}U^{\dagger}_{\xt_2}] \text{Tr}[t^{a'}U_{\xt_3}U^{\dagger}_{\xt_4}] \rangle
=
\begin{bmatrix}
 0 \\
 \frac{1}{2} \delta^{aa'} 
\end{bmatrix}
^{\mathrm{T}}
e^{ M_{2\times2}}
\begin{bmatrix}
1  \\
0
\end{bmatrix},
\end{equation}
which is exactly the correlator we needed for the 2-\(\alpha\) correlator
\begin{equation}
\begin{split}
&
\langle \alpha^{i,a}_{\xt} \alpha^{k,a'}_{\xt} \rangle
\\ &
=
\lim_{\substack{\xt_i \to \xt}}
\Bigg\{
-\frac{2}{g^2} \delta^{aa'} \partial ^i_{\textbf{x}_2} \partial ^k_{\textbf{x}_4}
\begin{bmatrix}
0 & 1
\end{bmatrix}
e^{M_{2\times2}}
\begin{bmatrix}
1 \\
0
\end{bmatrix}
\Bigg\}
\end{split}.
\end{equation}

For details for finding the transition matrix for a given basis see \cite{Blaizot:2004wv}, where it has been done for the 4-point Wilson line correlator. We present an algorithm for finding the matrix for the specific basis we have chosen here. Firstly let us give a name for the basis states for easier handling. Let us denote them with an \(S\) with 2-tuples as arguments, which tell what fundamental index is associated with which Wilson line and which Wilson lines are connected to each other. Each of these ``open dipoles'' are separated by a semicolon. For example we can write the first basis state as
\begin{equation}
\begin{split}
&
\mathcal{W}_1 = [U_{\xt_1}U^{\dagger}_{\xt_2}]_{k_1 k_2} [U_{\xt_3}U^{\dagger}_{\xt_4}]_{k_3 k_4} 
\\ &
\equiv S((\xt_1,k_1),(\xt_2,k_2);(\xt_3,k_3),(\xt_4,k_4)).
\end{split}
\end{equation}
The other basis state is just a permutation of the 2-tuples in such a way that we pair the Wilson lines with the daggered Wilson lines in a different way.

To find the transition matrix, we compute how the basis states evolve in one step of evolution i.e. when we have one fundamental color matrix contraction done in every possible way, omitting tadpoles for now. When organizing the computation in this way, it suffices to compute the evolution of one basis state as the other differs from it just by a permutation of 2-tuples. The evolution for the first basis state is
\begin{multline} \label{eq:4WLEvolution}
S((\xt_1,k_1),(\xt_2,k_2);(\xt_3,k_3),(\xt_4,k_4))
\\ 
\rightarrow
\Big \{ C_F \big [ L(\xt_1,\xt_2) + L(\xt_3,\xt_4) \big ] 
+
\frac{1}{2N} \Big [ L(\xt_1,\xt_3)+ L(\xt_2,\xt_4) 
\\ - L(\xt_1,\xt_4) - L(\xt_2,\xt_3) \Big ]
\Big\}
\\  \times
S((\xt_1,k_1),(\xt_2,k_2);(\xt_3,k_3),(\xt_4,k_4))
\\ 
+
\Big \{
\frac{1}{2} \big [ L(\xt_1,\xt_4) + L(\xt_2,\xt_3) - L(\xt_1,\xt_3) - L(\xt_2,\xt_4) \big ]
\Big \}
\\  \times
S((\xt_1,k_1),(\xt_4,k_4);(\xt_3,k_3),(\xt_2,k_2)).
\end{multline}
In the MV model, one would have
\begin{equation}
L(\xt,\yt) \equiv g^2 \mu^2 \int _{\zt} G(\xt - \zt)G(\yt - \zt),
\end{equation}
where the Green's function is the same as the one defined in \eqref{eq:GreensFunction}. Now by identifying the basis states and making the replacement \(\mathcal{W}_i \rightarrow e_i\), one has found a column vector of the transition matrix. If one does the evolution in equation \eqref{eq:4WLEvolution} for the i:th basis state, the resulting column vector is the i:th row of the transition matrix. Doing this for all the basis states, one finds the transition matrix, which is still missing the tadpole contributions. The tadpole contributions always combine with the ladder-type contributions in the same way forming the quantities
\begin{equation}
\Gamma(\xt_i,\xt_j) = L(\xt_i , \xt_i) + L( \xt_j , \xt_j) - 2L(\xt_i , \xt_j),
\end{equation}
which can be computed in the MV model using the definition of the \(L\)-function. The final transition matrix can then in practice be found with the replacement
\begin{equation} \label{eq:LToGammaReplacement}
L(\xt_i,\xt_j) \rightarrow - \frac{1}{2} \Gamma(\xt_i - \xt_j).
\end{equation}
As an example, the resulting transition matrix \(M_{2 \times 2}\), with this choice of basis, is
\begin{widetext}
\begin{equation}
M_{2 \times 2}
=
\begin{bmatrix}
- \frac{C_F}{2} \big [ \Gamma_{1,2} + \Gamma_{3,4} \big ] - \frac{1}{4N} \big [ \Gamma_{1,3} + \Gamma_{2,4} - \Gamma_{1,4} - \Gamma_{2,3} \big ] & -\frac{1}{4} \big [ \Gamma_{1,2} + \Gamma_{3,4} - \Gamma_{1,3} - \Gamma_{2,4} \big ] \\
-\frac{1}{4} \big [ \Gamma_{1,4} + \Gamma_{2,3} - \Gamma_{1,3} - \Gamma_{2,4} \big ] & -\frac{C_F}{2} \big [ \Gamma_{1,4} + \Gamma_{2,3} \big ] - \frac{1}{4N} \big [ \Gamma_{1,3} + \Gamma_{2,4} - \Gamma_{1,2} - \Gamma_{3,4} \big ]
\end{bmatrix}.
\end{equation}
\end{widetext}
Here we have used the short hand notation
\begin{equation}
\Gamma(\xt_i , \xt_j) \equiv \Gamma_{i,j}.
\end{equation}

In this work we use the GBW parametrization to describe the nucleus gluon fields. In Gaussian models the Wilson line correlators can be expressed in terms of the Wilson line dipole. One can thus match the \(\Gamma\)-function appearing in the dipole to the GBW dipole and use this \(\Gamma\) to obtain the higher point correlators. By computing the Wilson line dipole using the method we have used for the other correlators in this work, one can find that it can be written as
\begin{equation}
D_{\xt,\yt}= \exp \left\{ - \frac{1}{2} C_F  \Gamma(\xt - \yt) \right\}.
\end{equation}
On the other hand the GBW dipole can be written as \cite{GolecBiernat:1998js}
\begin{equation}
D^{GBW}_{\xt,\yt}= \exp \left\{ - \frac{Q_s^2}{4}(\xt - \yt)^2 \right\}.
\end{equation}
By matching these, one finds that the required transformation for expressing our results in the GBW model is
\begin{equation}
 \Gamma(\xt - \yt) \rightarrow \frac{Q_s^2}{2 C_F} (\xt - \yt)^2.
\end{equation}

The same calculation can be done for the fundamental 8-point Wilson line correlator that is required for the nucleus connected part of the energy 2-point function. In this case we have 24 states that mix with our correlator. Again, the correlators can be built from the basic building blocks \ref{eq:CorrelatorBuildingBlock} and can be found by pairing the Wilson lines with the daggered Wilson lines in every possible way. The correlator we wish to compute is
\begin{equation}
\langle  
  [U_{\xt_1} U^{\dagger}_{\xt_2}]_{k_1 k_2} 
  [U_{\xt_3} U^{\dagger}_{\xt_4}]_{k_3 k_4} 
  [U_{\xt_5} U^{\dagger}_{\xt_6}]_{k_5 k_6} 
  [U_{\xt_7} U^{\dagger}_{\xt_8}]_{k_7 k_8} 
\rangle.
\end{equation}

The exact choice of basis does not matter so we just set the state corresponding to the correlator to be the first basis state. Again, we write the basis states in a more handleable manner as
\begin{multline}
\mathcal{W}_1 =
  [U_{\xt_1} U^{\dagger}_{\xt_2}]_{k_1 k_2} 
  [U_{\xt_3} U^{\dagger}_{\xt_4}]_{k_3 k_4} 
  [U_{\xt_5} U^{\dagger}_{\xt_6}]_{k_5 k_6} 
  [U_{\xt_7} U^{\dagger}_{\xt_8}]_{k_7 k_8} 
  \\ 
  \equiv
  S((\xt_1,k_1),(\xt_2,k_2);(\xt_3,k_3),(\xt_4,k_4);
  \\ 
  (\xt_5,k_5),(\xt_6,k_6);(\xt_7,k_7),(\xt_8,k_8)).
\end{multline}
As with the 4-point Wilson line correlator, the other basis states can be found by permuting the 2-tuples acting as arguments for \(S\). Now we need to compute how the basis states evolve. Again, it suffices to only compute this for one basis state with others being found through permutations of coordinates and indices of the result. Let us evolve the first basis state. One step in the evolution reads
\begin{widetext}
\begin{equation}
\begin{split}
&
\!\!\!\!\!\!\!\!\!\!\!\!\!\!\!
S((\xt_1,k_1),(\xt_2,k_2);(\xt_3,k_3),(\xt_4,k_4);(\xt_5,k_5),(\xt_6,k_6);(\xt_7,k_7),(\xt_8,k_8))
\\ &
\!\!\!\!\!\!\!\!\!\!
\rightarrow
\Big \{ C_F \big [ L_{1,2} + L_{3,4} + L_{5,6} + L_{7,8} \big ]
\\ &
+ \frac{1}{2N} \big [ L_{1,3} - L_{1,4} + L_{1,5} - L_{1,6} + L_{1,7} - L_{1,8} - L_{2,3} + L_{2,4}
\\ &
 - L_{2,5} + L_{2,6} - L_{2,7} + L_{2,8} + L_{3,5} - L_{3,6} + L_{3,7} - L_{3,8}
\\ & 
  - L_{4,5} + L_{4,6} - L_{4,7} + L_{4,8} + L_{5,7} - L_{5,8} - L_{6,7} + L_{6,8} \big ]
 \Big \}
\\ & 
\quad\quad\quad\quad
\times
S((\xt_1,k_1),(\xt_2,k_2);(\xt_3,k_3),(\xt_4,k_4);(\xt_5,k_5),(\xt_6,k_6);(\xt_7,k_7),(\xt_8,k_8))
\\ &
+
\frac{1}{2} \big [ -L_{1,3} + L_{1,4} + L_{2,3} - L_{2,4} \big ]
S((\xt_1,k_1),(\xt_4,k_4);(\xt_3,k_3),(\xt_2,k_2);(\xt_5,k_5),(\xt_6,k_6);(\xt_7,k_7),(\xt_8,k_8))
\\ &
+
\frac{1}{2} \big [ -L_{1,5} + L_{1,6} + L_{2,5} - L_{2,6} \big ]
S((\xt_1,k_1),(\xt_6,k_6);(\xt_3,k_3),(\xt_4,k_4);(\xt_5,k_5),(\xt_2,k_2);(\xt_7,k_7),(\xt_8,k_8))
\\ &
+
\frac{1}{2} \big [ -L_{1,7} + L_{1,8} + L_{2,7} - L_{2,8} \big ]
S((\xt_1,k_1),(\xt_8,k_8);(\xt_3,k_3),(\xt_4,k_4);(\xt_5,k_5),(\xt_6,k_6);(\xt_7,k_7),(\xt_2,k_2))
\\ &
+
\frac{1}{2} \big [ -L_{3,5} + L_{3,6} + L_{4,5} - L_{4,6} \big ]
S((\xt_1,k_1),(\xt_2,k_2);(\xt_3,k_3),(\xt_6,k_6);(\xt_5,k_5),(\xt_4,k_4);(\xt_7,k_7),(\xt_8,k_8))
\\ &
+
\frac{1}{2} \big [ -L_{3,7} + L_{3,8} + L_{4,7} - L_{4,8} \big ]
S((\xt_1,k_1),(\xt_2,k_2);(\xt_3,k_3),(\xt_8,k_8);(\xt_5,k_5),(\xt_6,k_6);(\xt_7,k_7),(\xt_4,k_4))
\\ &
+
\frac{1}{2} \big [ -L_{5,7} + L_{5,8} + L_{6,7} - L_{6,8} \big ]
S((\xt_1,k_1),(\xt_2,k_2);(\xt_3,k_3),(\xt_4,k_4);(\xt_5,k_5),(\xt_8,k_8);(\xt_7,k_7),(\xt_6,k_6))
\end{split}
\end{equation}

The evolution of other basis states can be found by permuting the 2-tuples to match them. After this, we can again do the replacement \(\mathcal{W}_i \rightarrow e_i\) and use the result as the i:th column of the transition matrix \(M_{24 \times 24}\). Again, the tadpole contribution can be added to the matrix by the replacement in \eq\nr{eq:LToGammaReplacement}.

Now we have all we need to compute the 4-\(\alpha\) correlator we needed for the nucleus connected contribution
\begin{equation}
\langle \alpha^{i,a}_{\xt} \alpha^{k,c}_{\xt} \alpha^{i',a'}_{\yt} \alpha^{k',c'}_{\yt} \rangle
 =
\lim_{\substack{\xt_{1,2,3,4} \to \xt \\ \xt_{5,6,7,8} \to \yt}}
\left\{
\frac{16}{g^4}
\partial^i_{\xt_2} \partial^k_{\xt_4} \partial^{i'}_{\xt_6} \partial^{k'}_{\xt_8}
\begin{bmatrix}
\delta^{k_1 k_2} \delta^{k_3 k_4}  \delta^{k_5 k_6} \delta^{k_7 k_8} \\ \vdots
\end{bmatrix}
^{\mathrm{T}}
e^{M_{24\times24}}
\begin{bmatrix}
t^a_{k_2 k_1} t^c_{k_4 k_3} t^{a'}_{k_6 k_5} t^{c'}_{k_8 k_7} \\
0 \\
\vdots \\
0
\end{bmatrix}
\right\}.
\end{equation}
\end{widetext}

As the evaluation of the matrix exponential is the most cumbersome part of computing this correlator, one can use the identity
\begin{equation}
\partial_x e^{M(x)} = \int_0^1 e^{tM}[\partial_x M]e^{(1-t)M} dt,
\end{equation}
to simplify the calculation. One can take the derivatives using the identity and then take the coordinate limits in the exponents, which causes some of the \(\Gamma\)-functions to vanish.

\begin{widetext}
The expression we obtain for the color contracted nucleus correlator is
\begin{equation}\label{eq:bigeq}
\begin{split}
&
\langle \alpha_{\xt}^{i,a} \alpha_{\xt}^{j,a} \alpha_{\yt}^{k,c} \alpha_{\yt}^{l,c} \rangle
\\ & =
\lim_{\substack{\xt_{1,2,3,4} \rightarrow \xt \\ \xt_{5,6,7,8} \rightarrow \yt}}
\Bigg\{
\frac{1}{g^4 N^4 (C_A - 1)^2 (C_A + 1)^2 \Gamma^2 _{\xt , \yt}}
C_A^2 e^{-\frac{1}{2}(9 C_A +2)\Gamma_{\xt , \yt}}
\\ & \quad \quad\quad  \times
\Bigg[ -  C_A (C_A^2 -1)^2 \Gamma_{\xt , \yt} e^{(4 C_A +1)\Gamma_{\xt , \yt}} \Big ( -2 C_A C_F  \Big ( e^{\frac{1}{2}  C_A \Gamma_{\xt , \yt}} + 1 \Big ) + C_A^2 -1 \Big )
\Big ( \Gamma_{\xt_2 , \xt_8}^{i,l} \Gamma_{\xt_4 , \xt_6}^{j,k} + \Gamma_{\xt_2 , \xt_6}^{i,k} \Gamma_{\xt_4 , \xt_8}^{j,l} \Big )
\\ &
 \quad \quad \quad\quad \quad 
- \Bigg( C_A^6 \Big ( C_F \Big ( e^{2  \Gamma_{\xt , \yt}}-1 \Big ) e^{\frac{7}{2} C_A \Gamma_{\xt , \yt}} + 2 \Big ( e^{( 4 C_A + 1) \Gamma_{\xt , \yt}} - e^{\frac{1}{2} (9C_A +2)\Gamma_{\xt , \yt}} \Big ) \Big )
\\ &
 \quad \quad \quad\quad \quad\quad 
 + 2 C_A^5 C_F e^{\frac{7}{2}  C_A \Gamma_{\xt , \yt}} \Big ( 4 e^{ (C_A +1)\Gamma_{\xt , \yt}} - 6 e^{\frac{1}{2} (C_A + 2)\Gamma_{\xt , \yt}}   - C_F e^{2  \Gamma_{\xt , \yt}}  + e^{2 \Gamma_{\xt , \yt}} + C_F + 1 \Big )
\\ &
 \quad \quad \quad\quad \quad\quad 
+ C_A^4 \Big ( - 4 C_F^2 \Big ( e^{\frac{7}{2}  C_A \Gamma_{\xt , \yt}} -2 e^{ (4C_A +1)\Gamma_{\xt , \yt}} + e^{\frac{1}{2}  (7C_A + 4)\Gamma_{\xt , \yt}} \Big ) - 6e^{(4C_A + 1)\Gamma_{\xt , \yt}}+6e^{\frac{1}{2}  (9C_A + 2)\Gamma_{\xt , \yt}} \Big )
\\ &
 \quad \quad \quad\quad \quad\quad 
- 2 C_A^3 C_F e^{\frac{7}{2}  C_A \Gamma_{\xt , \yt}} \Big ( 10e^{ (C_A + 1)\Gamma_{\xt , \yt}} - 12e^{\frac{1}{2}  (C_A + 2)\Gamma_{\xt , \yt}} + C_F \Big ( e^{2 \Gamma_{\xt , \yt}} - 1 \Big ) + e^{2  \Gamma_{\xt , \yt}} +1 \Big )
\\ & 
 \quad \quad \quad\quad \quad\quad 
+ C_A^2 \Big ( -8C_F^2 \Big ( e^{ (4C_A +1)\Gamma_{\xt , \yt}} - e^{\frac{1}{2} (9 C_A +2)\Gamma_{\xt , \yt}} \Big ) -C_F \Big ( e^{2 \Gamma_{\xt , \yt}}-1 \Big ) e^{\frac{7}{2} C_A \Gamma_{\xt , \yt}}
+6 \Big ( e^{ (4 C_A +1)\Gamma_{\xt , \yt}} - e^{\frac{1}{2} (9C_A +2)\Gamma_{\xt , \yt}} \Big ) \Big )
\\& \quad \quad \quad\quad \quad\quad 
-12C_A C_F \Big ( e^{(4C_A +1)\Gamma_{\xt , \yt}} - e^{\frac{1}{2} (9C_A +2)\Gamma_{\xt , \yt}} \Big )
- 2 e^{ (4C_A + 1)\Gamma_{\xt , \yt}} + 2e^{\frac{1}{2}  (9C_A +2)\Gamma_{\xt , \yt}} \Bigg) 
\Big ( \Gamma_{\xt_2 , \xt_8}^{i,l}\Gamma_{\xt_4 , \xt_6}^{j,k} + \Gamma_{\xt_2 , \xt_6}^{i,k}\Gamma_{\xt_4 , \xt_8}^{j,l} \Big )
\\ &
 \quad \quad \quad\quad \quad
+  C_A^4 (C_A^2 -1)^2 C_F^2 \Gamma_{\xt , \yt}^2 e^{\frac{1}{2} (9 C_A + 2)\Gamma_{\xt , \yt}} \Gamma_{\xt_2 , \xt_4}^{i,j}\Gamma_{\xt_6 , \xt_8}^{k,l} \Bigg ] \Bigg\}.
\end{split}
\end{equation}
\end{widetext}
We have assumed that the model used has the property
\begin{equation}
\lim _{\xt \rightarrow \yt} \partial ^i _{\xt } \Gamma (\xt , \yt) = 0,
\end{equation}
but did not make further assumptions of the model used for the infinite, homogenous nucleus. The correlator 
in \eq\nr{eq:bigeq} is the one needed for the contraction of the proton disconnected part with the nucleus connected part of the energy density 2-point function. The superscripts on the \(\Gamma\)-functions indicate derivatives. \(i\) is with respect to \(\xt _2\), \(j\) is with respect to \(\xt _4\), \(k\) is with respect to \(\xt _6\) and \(l\) is with respect to \(\xt _8\). So for example
\begin{equation}
\Gamma_{\xt _2 , \xt _4}^{i,j} = \partial ^i _{\xt _2} \partial ^j _{\xt _4} \Gamma (\xt _2 , \xt _4).
\end{equation}
After taking all the derivatives, one has to take the limits \(\xt _{1,2,3,4} \rightarrow \xt\) and \(\xt _{5,6,7,8} \rightarrow \yt\). We used the FeynCalc package to handle the required color algebra for this correlator~\cite{Shtabovenko:2016sxi,Shtabovenko:2020gxv,Mertig:1990an}.
\begin{figure}[h!]
    \centering
    \subfloat[\(n=2\)]{
    \includegraphics[width=\columnwidth]{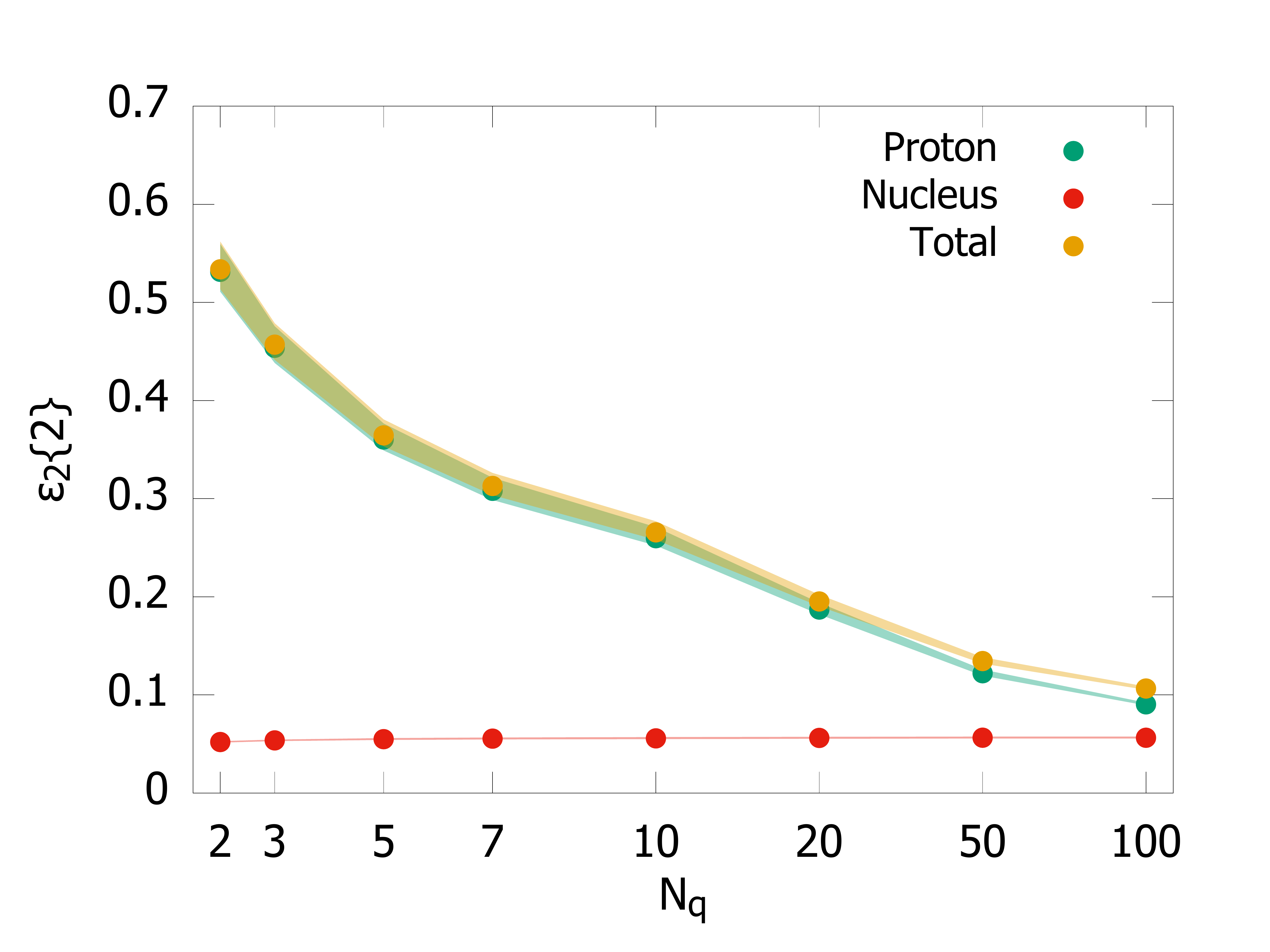}
    }
    \vspace{-2em}
    \subfloat[\(n=3\)]{
    \includegraphics[width=\columnwidth]{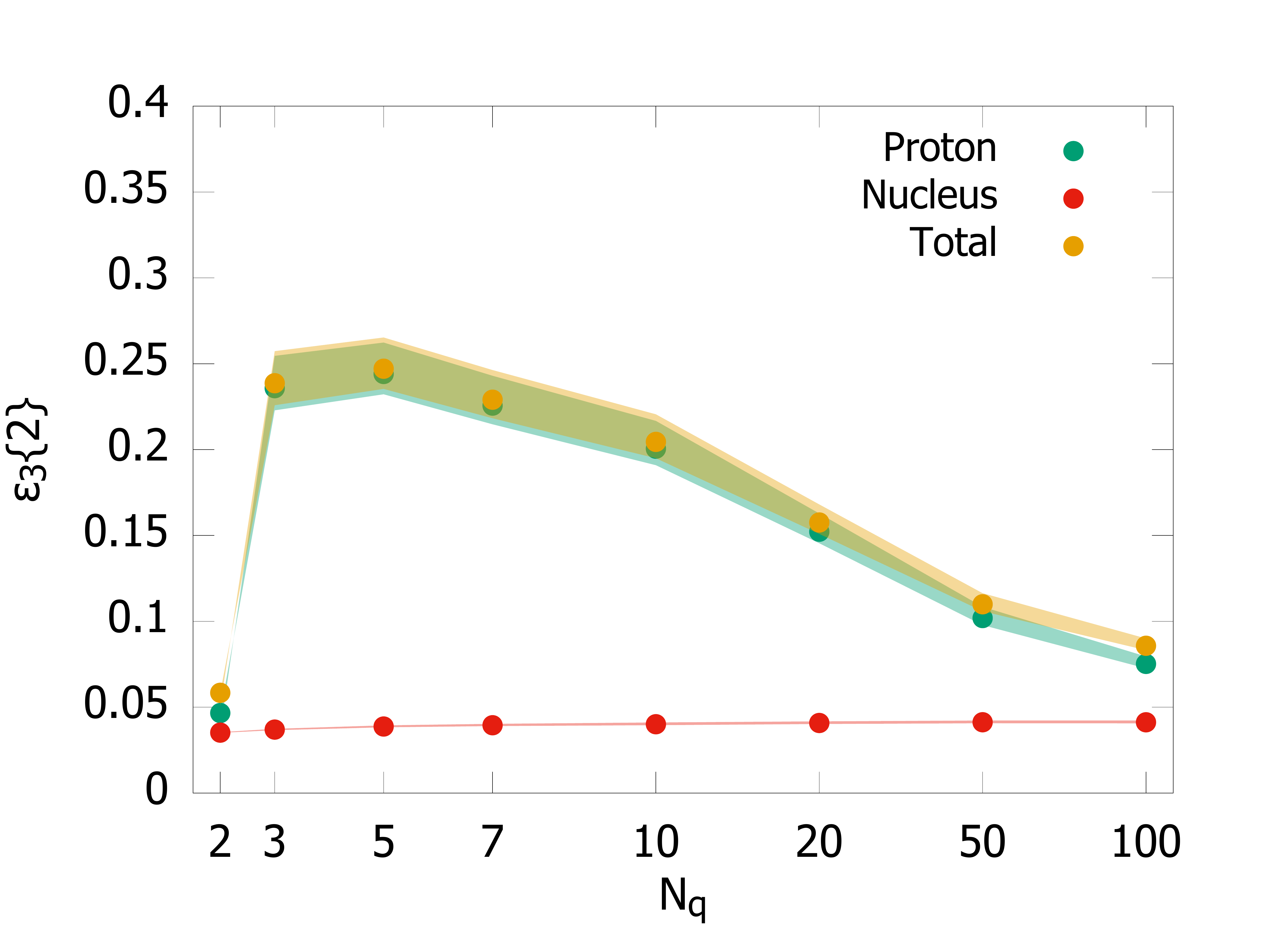}
    }
    \vspace{-2em}
    \subfloat[\(n=4\)]{
    \includegraphics[width=\columnwidth]{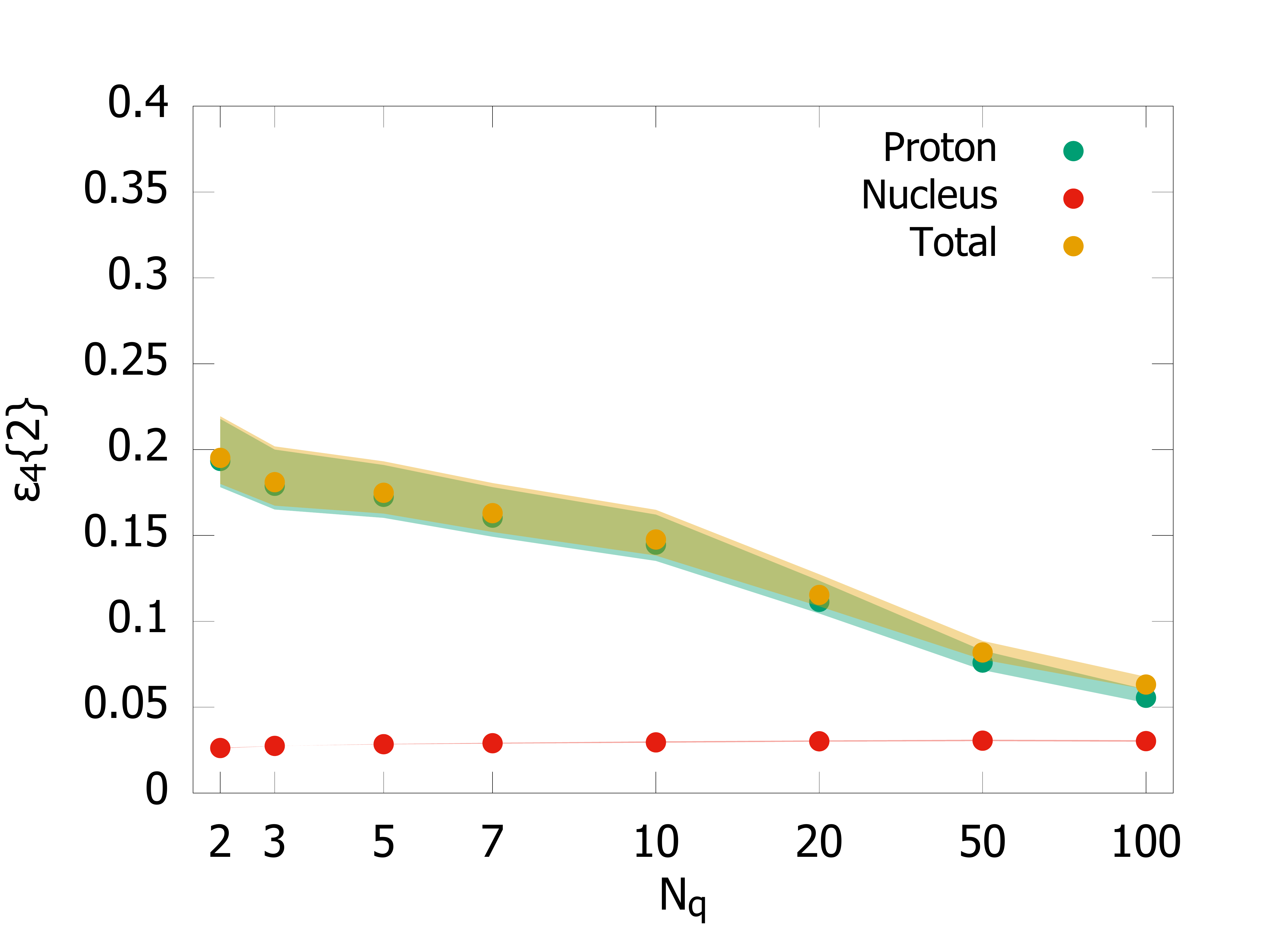}
    }
\caption{Eccentricities $\varepsilon_{n}\{2\}$ as functions of the number of hot spots \(N_q\) for \(n=2,3,4\). The bands are obtained by varying the UV cutoff \(C_0\) by \(\pm 50\%\).}
\label{fig:n234EllipticityVsNqC0}
\end{figure}

\section{Some additional results}
\label{app:numerics}

Our model required two different regularization parameters, one for the UV and one for the IR divergences. When presenting results for the eccentricities, we have used mass variation to obtain confidence bands for our results, as that was the major source of uncertainty among these two parameters. In figure \ref{fig:n234EllipticityVsNqC0} we present some example plots that show that the UV cutoff \((C_0)\) dependence of the eccentricities is indeed much smaller than their dependence on the IR mass regulator \((m)\).

\begin{figure}[h!]
    \centering
    \includegraphics[width=\columnwidth]{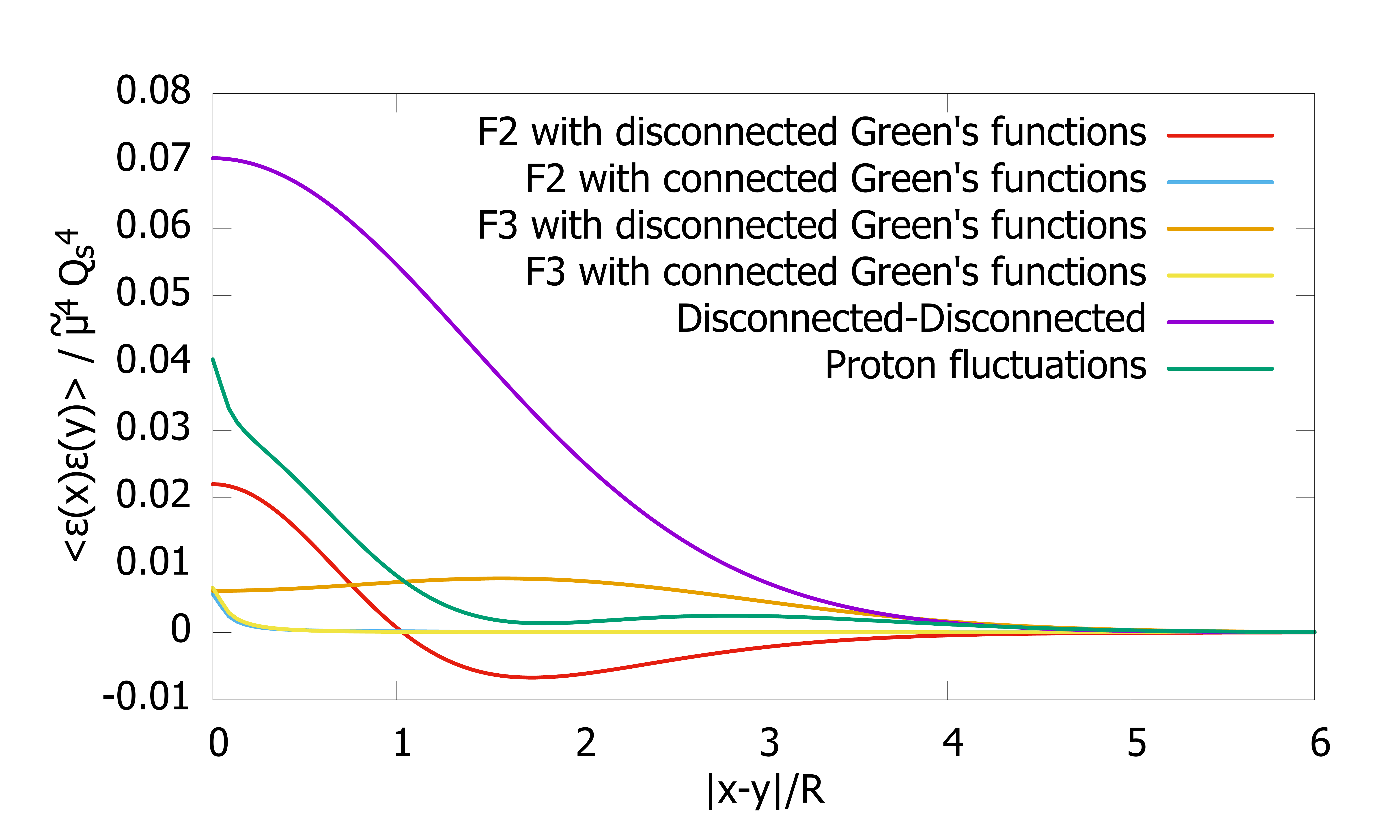}
    \caption{Breakup of different contributions to the proton fluctuation contribution to the two-point function of the energy density $\langle \varepsilon(\xt)\varepsilon(\yt)\rangle$. }
    \label{fig:ProtonTwoPointContributionsProportionalDC}
\end{figure}

\begin{figure}[h!]
    \centering
    \includegraphics[width=\columnwidth]{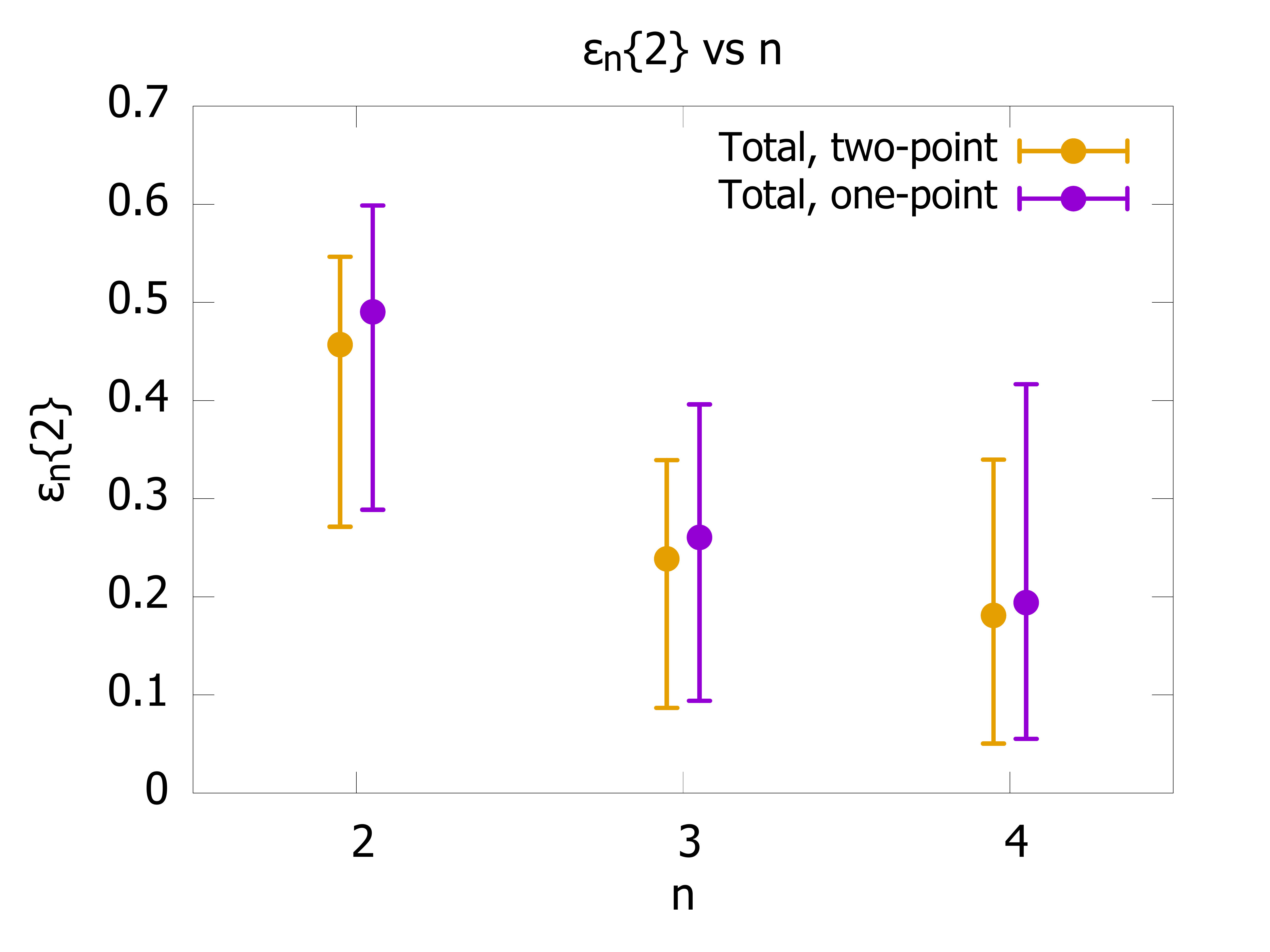}
    \caption{Comparison of two different definitions of the eccentricity $\varepsilon_{n}\{2\}$ and $\varepsilon_{n}^{'}\{2\}_{\rm Approx}$ corresponding to normalizing the eccentricity by energy density two-point function or the square of the one-point function. Numerical results are obtained for \(N_q=3\) and error bars indicate the sensitivity to variations of the IR regulator \(m\) by \(\pm 50\%\) . }
    \label{fig:Nq3EllipticityFullVsApprox}
\end{figure}

\FloatBarrier

In figure \ref{fig:ProtonTwoPointContributionsProportionalDC} we have plotted the different parts of the proton connected contribution to the 2-point function of the energy density to explain the emergence of the second peak in the right-hand side plot in figure \ref{fig:Nq3TwoPointMassLogPlots}.
Different contributions to the proton fluctuations can  be separated into the following contributions

\begin{widetext}
\begin{equation} \label{eq:protonConnectedPartsSeparated}
\begin{split}
&
\langle \varepsilon(\textbf{x})\varepsilon(\textbf{y}) \rangle _{\mathrm{DC,C}}
\\
&
=
\underbrace{
\frac{Q_s^4 N^2}{4 C_F^2}
 \int \ud^2 \at \ud^2 \bt N_q \Big[ F_2(\at, \bt, \Bt) - F_1(\at, \Bt)F_1(\bt, \Bt) \Big]
(N^2-1)^2 
\textbf{G}^i_{\xt}(\xt-\at)
\textbf{G}^i_{\xt}(\xt-\at)
\textbf{G}^j_{\yt}(\yt-\bt)
\textbf{G}^j_{\yt}(\yt-\bt)
}_{\text{$F_2$ with disconnected Green's functions}}
\\ &
+
\underbrace{
\frac{Q_s^4 N^2}{4 C_F^2}
 \int \ud^2 \at \ud^2 \bt N_q F_2(\at, \bt, \Bt)
2(N^2-1)
\textbf{G}^i_{\xt}(\xt-\at)
\textbf{G}^i_{\xt}(\xt-\bt)
\textbf{G}^j_{\yt}(\yt-\at)
\textbf{G}^j_{\yt}(\yt-\bt)
}_{\text{$F_2$ with connected Green's functions}}
\\ &
+
\underbrace{
\frac{Q_s^4 N^2}{4 C_F^2}
 \int \ud^2 \at \ud^2 \bt N_q(N_q-1) \Big[ F_3(\at, \bt, \Bt) - F_1(\at, \Bt)F_1(\bt, \Bt) \Big]
(N^2-1)^2 
\textbf{G}^i_{\xt}(\xt-\at)
\textbf{G}^i_{\xt}(\xt-\at)
\textbf{G}^j_{\yt}(\yt-\bt)
\textbf{G}^j_{\yt}(\yt-\bt)
}_{\text{$F_3$ with disconnected Green's functions}}
\\ &
+
\underbrace{
\frac{Q_s^4 N^2}{4 C_F^2}
 \int \ud^2 \at \ud^2 \bt N_q(N_q-1)F_3(\at, \bt, \Bt)
2(N^2-1)
\textbf{G}^i_{\xt}(\xt-\at)
\textbf{G}^i_{\xt}(\xt-\bt)
\textbf{G}^j_{\yt}(\yt-\at)
\textbf{G}^j_{\yt}(\yt-\bt)
}_{\text{$F_3$ with connected Green's functions}}
\end{split}
\end{equation}
\end{widetext}
where the disconnected contribution is subtracted proportionally from the terms with the disconnected Green's functions as it seems like the natural way to distribute it between the two contributions. In this way all of the disconnected contribution is subtracted from the \(F_2\) term when \(N_q=1\) i.e. when the \(F_3\) term vanishes. Then an increasing fraction of the disconnected contribution is subtracted from the \(F_3\) term as it grows with a growing \(N_q\).  By inspecting Fig.~\ref{fig:ProtonTwoPointContributionsProportionalDC} one observes that the first peak in the proton fluctuations is dominated by the contributions from a single hot spot $(F_2)$, while the second peak emerges when the contribution from two hot spots $(F_3)$ becomes dominant.

In figure \ref{fig:Nq3EllipticityFullVsApprox} we show how our definition of eccentricity \eqref{eq:EccTwoPointDenom} differs from the approximation in equation \eqref{eq:EccApprox}. In the parameter space we have been exploring in this paper, the results of these two definitions do not differ much. However, as one moves to the parts of the parameter space, where fluctuations are large, for example when the hot spots are very localized, one starts to see that the the result obtained from the two different definitions start to differ more. In the extreme case of energy density being a collection of Dirac delta-like hot spots as in Appendix~\ref{app:delta}, the approximation \eqref{eq:EccApprox} is not bounded by one unlike the definition \eqref{eq:EccTwoPointDenom} that we use throughout this paper.

%%%%%%%%%%%%%%%%%%%%%%%%%%%%%%%%%%
\section{Pointlike energy density model}
\label{app:delta}

We consider the limiting behaviour for our model by taking the energy density to be a collection of Dirac delta hot spots of extremely localized gluon fields. This should be the limit of our model if we formally took the limits of zero hot spot size \((r \rightarrow 0)\) and infinite mass \((m \rightarrow \infty)\).  The expression for the energy density is now
\begin{equation}
\varepsilon (\xt) = \varepsilon_0 \sum ^{N_q}_{i=1} \delta^2 (\xt - \bt _i),
\end{equation}
where, as before, $\bt_i$ denotes the position of the hot spots and \(\varepsilon_0\) is a dimensionful constant whose exact value is of no importance as it cancels out in the end. 

Now having the expression for the energy density, we can compute its one and two point functions  using the hot spot averaging procedure yielding
\begin{multline}
\langle \varepsilon (\xt) \rangle =
\left( \frac{\varepsilon_0}{2\pi R^2} \right) \left( \frac{N_q^2}{N_q-1} \right)
\\  \times
\exp \Big \{ -\frac{1}{2R^2} 
\Big ( \frac{N_q}{N_q-1} 
\Big ) (\xt-\Bt)^2 \Big \},
\end{multline}
which is valid for \(N_q \geq 2\), and
\begin{equation}
\begin{split}
&
\langle \varepsilon (\xt) \varepsilon (\yt) \rangle
=
\left( \frac{\varepsilon_0^2}{2 \pi R^2} \right) \left( \frac{N_q^2}{N_q-1} \right) \delta^2(\xt - \yt)
\\ & 
\quad\quad 
\times
\exp \Big \{ -\frac{1}{2R^2} 
\Big ( \frac{N_q}{N_q-1} 
\Big ) (\xt-\Bt)^2 \Big \}
\\ & +
\Big ( \frac{\varepsilon_0}{2 \pi R^2} \Big )^2 \Big ( \frac{N_q^2(N_q-1)}{N_q-2} \Big )
\exp \Big \{ -\frac{1}{4R^2} (\xt - \yt)^2
\\ & 
\quad\quad
-\frac{1}{4R^2} 
\Big ( \frac{N_q}{N_q-2}
\Big )(\xt + \yt -2\Bt)^2
\Big \}
\end{split}
\end{equation}
which is valid for \(N_q \geq 3\). 

Based on the above results of the one and two-point correlation functions of the energy density, we now we go on and compute eccentricities $\varepsilon_{n}\{2\}$ defined in \eq\nr{eq:EccTwoPointDenom} in this model. For the numerator we need to compute the weighted integral of the two-point function:
\begin{widetext}
\begin{multline} \label{eq:deltaTwoPointNumerator}
\int \ud^2\xt \ud^2\yt |\xt-\Bt|^n|\yt-\Bt|^n \cos (n \theta_{\xt-\Bt} - n \theta_{\yt-\Bt}) \langle \varepsilon (\xt) \varepsilon (\yt) \rangle
=
\\ 
\varepsilon _0^2 R^{2 n} (-2)^n (N_q-1) N_q^{1-n} n \Gamma (n)
%\\
+ 
\varepsilon _0^2 R^{2 n} N_q \Big ( 2-\frac{2}{N_q} \Big )^n n\Gamma (n),
\end{multline}
where here and in subsequent equations \(\Gamma(n)\) is the usual gamma function. We further need the denominator integral for the two-point function, and to be able to compare the definitions of eccentricity, the integral of the energy density one-point function squared:
\begin{multline}
\int \ud^2\xt \ud^2\yt |\xt-\Bt|^n|\yt-\Bt|^n \langle \varepsilon (\xt) \varepsilon (\yt) \rangle
=
\\ 
\varepsilon_0^2 R^{2 n} 2^n N_q^2 \Big ( \frac{N_q-2}{N_q-1} \Big ) ^{n+1} \Gamma \Big ( \frac{n}{2}+1 \Big ) ^2 
% \\  \times
{}_2F_1 \Big ( \frac{n+2}{2},\frac{n+2}{2};1;\frac{1}{(N_q-1)^2} \Big )
%\\
+ \varepsilon _0^2 R^{2 n} N_q \Big ( 2-\frac{2}{N_q} \Big ) ^n n \Gamma (n)
\end{multline}
and
\begin{equation}
\int \ud^2\xt \ud^2\yt |\xt-\Bt|^n|\yt-\Bt|^n \langle \varepsilon (\xt) \rangle \langle \varepsilon (\yt) \rangle
% \\ 
=
\varepsilon _0^2 R^{2 n} N_q^2 \Big ( 2-\frac{2}{N_q} \Big ) ^n  \Gamma \Big ( \frac{n}{2}+1 \Big ) ^2 \;.
\end{equation}
By combining the results, the eccentricities in the point like energy density model are then given by
\begin{equation}
\varepsilon_{n}\{2\}=
\sqrt{\frac{(-2)^n (N_q-1) N_q^{1-n} n \Gamma (n)
+ 
N_q \Big ( 2-\frac{2}{N_q} \Big )^n n\Gamma (n)}{ 2^n N_q^2 \Big ( \frac{N_q-2}{N_q-1} \Big ) ^{n+1} \Gamma \Big ( \frac{n}{2}+1 \Big ) ^2 
 \times
{}_2F_1 \Big ( \frac{n+2}{2},\frac{n+2}{2};1;\frac{1}{(N_q-1)^2} \Big )
 +
N_q \Big ( 2-\frac{2}{N_q} \Big ) ^n n \Gamma (n)}}
\end{equation}
\end{widetext}
for $N_{q}\geq 3$ and we expect these values to give the absolute upper limit for the eccentricities in our hot spot model. 

By looking at the results in Fig.~\ref{fig:n234EllipticityVsNqMass} one can notice that in the point like energy density model, the \(n=3\) eccentricity is smaller than the \(n=4\) eccentricity. This is due to the fact that we have fixed the center of the proton and thus there is a preference for the hot spots to be ``back-to-back''. This effect shows up as the first term in~\nr{eq:deltaTwoPointNumerator} being proportional to $(-1)^n$. Now because of this, the relative angle between \(\xt\) and \(\yt\), the two points we measure the energy density at, is going to preferably be close to \(\pi\). Thus the \(\cos(n \theta _{\xt} - n \theta _{\yt})\) is preferably going to be \(\cos(n \pi)\), which gives a positive sign for the \(n\) even eccentricities and a negative sign for the \(n\) odd eccentricities as seen in \eqref{eq:deltaTwoPointNumerator}. The other term is due to the gluon fields being taken from the same pointlike source, for which the cosine always gives a 1. A hint of this same behavior can be observed in our actual model when considering the upper error bands of our eccentricities with a small number of hot spots. This makes sense as the upper bound corresponds to a larger mass which results in a more localized gluon field originating from a hot spot.

\bibliographystyle{JHEP-2modlong}
\bibliography{refs}

\end{document}